\documentstyle[12pt,epsfig]{article}
 
\newcommand{\agt}{\,\rlap{\lower 3.5 pt \hbox{$\mathchar \sim$}} \raise 1pt
 \hbox {$>$}\,}
\newcommand{\alt}{\,\rlap{\lower 3.5 pt \hbox{$\mathchar \sim$}} \raise 1pt
 \hbox {$<$}\,}

\catcode`@=11
\newcount\@tempcntc
\def\@citex[#1]#2{\if@filesw\immediate\write\@auxout{\string\citation{#2}}\fi
  \@tempcnta\z@\@tempcntb\m@ne\def\@citea{}\@cite{\@for\@citeb:=#2\do
    {\@ifundefined
       {b@\@citeb}{\@citeo\@tempcntb\m@ne\@citea\def\@citea{,}{\bf ?}\@warning
       {Citation `\@citeb' on page \thepage \space undefined}}%
    {\setbox\z@\hbox{\global\@tempcntc0\csname b@\@citeb\endcsname\relax}%
     \ifnum\@tempcntc=\z@ \@citeo\@tempcntb\m@ne
       \@citea\def\@citea{,}\hbox{\csname b@\@citeb\endcsname}%
     \else
      \advance\@tempcntb\@ne
      \ifnum\@tempcntb=\@tempcntc
      \else\advance\@tempcntb\m@ne\@citeo
      \@tempcnta\@tempcntc\@tempcntb\@tempcntc\fi\fi}}\@citeo}{#1}}
\def\@citeo{\ifnum\@tempcnta>\@tempcntb\else\@citea\def\@citea{,}%
  \ifnum\@tempcnta=\@tempcntb\the\@tempcnta\else
   {\advance\@tempcnta\@ne\ifnum\@tempcnta=\@tempcntb \else \def\@citea{--}\fi
    \advance\@tempcnta\m@ne\the\@tempcnta\@citea\the\@tempcntb}\fi\fi}
\catcode`@=12

\begin{document}

\title{\vskip-3cm{\baselineskip14pt
\centerline{\normalsize DESY 00-086\hfill ISSN 0418-9833}
\centerline{\normalsize MPI/PhT/2000-10\hfill March 2000 }
}
\vskip1.5cm
Fragmentation Functions for Pions, Kaons, and Protons at Next-to-Leading Order
}
\author{{\sc B.A. Kniehl,$^1$ G. Kramer,$^1$ B. P\"otter$^2$}\\
{\normalsize $^1$ II. Institut f\"ur Theoretische Physik, Universit\"at
Hamburg,}\\
{\normalsize Luruper Chaussee 149, 22761 Hamburg, Germany}\\
{\normalsize $^2$ Max-Planck-Institut f\"ur Physik
(Werner-Heisenberg-Institut),}\\
{\normalsize F\"ohringer Ring 6, 80805 Munich, Germany}}

\date{}

\maketitle

\thispagestyle{empty}

\begin{abstract}
We present new sets of fragmentation functions for charged pions, charged 
kaons, and protons, both at the leading and next-to-leading orders.
They are fitted to the scaled-momentum distributions of these hadrons measured
in $e^+e^-$ annihilation on the $Z$-boson resonance at CERN LEP1 and SLAC SLC.
These data partly come as light-, charm-, bottom-quark-enriched and gluon-jet
samples, which allows us to treat all partons independently, after imposing 
the SU(2) flavour symmetry relations.
In order to gain sensitivity to the scaling violation in fragmentation, we
also include data from SLAC PEP, with center-of-mass energy $\sqrt s=29$~GeV,
in our fits.
This allows us to also determine the strong-coupling constant, with a 
competitive error.
LEP1 data on the longitudinal cross section as well as DESY DORIS and PETRA
data at lower energies nicely agree with theoretical predictions based on our
fragmentation functions.
\medskip

\end{abstract}

\newpage

\section{Introduction}

The inclusive production of a single charged hadron, $h$, in the annihilation
process
\begin{equation}
e^+e^-\to(\gamma,Z)\to h+X,
\label{process}
\end{equation}
has been measured at many different $e^+e^-$ colliders over a wide range of
center-of-mass (CM) energies, $\sqrt{s}$, between 3 and 183~GeV \cite{1,A}.
Here, $h$ may either refer to a specific hadron species, {\it e.g.},
pion, kaon, or proton, or to the sum over all charged-hadron species.
A large amount of precise data has now become available from various
experimental collaborations at CERN LEP1 and SLAC SLC, who started taking data
several years ago.
The process~(\ref{process}) is particularly suitable in order to study the
fragmentation of quarks and gluons into hadrons.
The information contained in fixed-target, hadron-collider, and
$ep$-scattering data is less useful, since it is obscured by theoretical
uncertainties from the parton distribution functions and the choice of
factorization scales connected with the initial states.

The partonic cross sections pertinent to process~(\ref{process}) can be
entirely calculated in perturbative QCD with no additional input, except for
the strong-coupling constant $\alpha_s$.
They are known at next-to-leading order (NLO) \cite{3} and even at
next-to-next-to-leading order (NNLO) \cite{rn96}.
The subsequent transition of the partons into hadrons takes place at an energy
scale of the order of 1~GeV and can, therefore, not be treated in perturbation
theory.
Instead, the hadronization of the partons is described by fragmentation
functions (FFs), $D_a^h(x,Q^2)$.
Their values correspond to the probability that the parton $a$, which is
produced at short distance, of order $1/Q$, fragments into the hadron $h$
carrying the fraction $x$ of the momentum of $a$.
In the case of process~(\ref{process}), $Q^2$ is typically of order $s$.
Given their $x$ dependence at some scale $Q_0^2$, the evolution of the FFs
with $Q^2$ may be computed perturbatively.
At present, the evolution equations are only known through NLO.
Consistency requires that we do not include the NNLO corrections to the
partonic cross sections \cite{rn96} in our analysis of process~(\ref{process})
via unpolarized photons and $Z$ bosons.

The theoretical analysis of the fragmentation process in $e^+e^-$ annihilation
is of interest for several reasons.
First, it allows us to test QCD quantitatively within one experiment observing
hadrons at different values of $\sqrt{s}$ and to determine the strong-coupling
constant, $\alpha_s(Q)$, to be compared with determinations from other
observables and/or processes.
Second, since according to the factorization theorem the FFs are independent
of the process in which they have been determined, they can be used for
quantitative predictions of inclusive single hadron cross sections in other
processes, like $p\bar{p}$, $ep$, $\gamma p$, and $\gamma\gamma$ scattering.
Third, the exact knowledge of the FFs, in particular for identified hadrons,
may help to elucidate the fundamental features of hadronization and to
constrain models used for the calculation of complete final states in
various high-energy reactions.

After the pioneering leading-order (LO) analyses of pion, kaon \cite{fie,bar},
and charmed-meson \cite{seh} FFs in the late 70s, there had long been no 
progress on the theoretical side of this field.
In the mid 90's, NLO FF sets for $\pi^0$ \cite{chi}, $\pi^\pm$, $K^\pm$, and
$\eta$ mesons \cite{gre} have been constructed through fits to data of
$e^+e^-$ annihilation, mostly generated with Monte Carlo event generators.

In 1994/95, two of us, together with Binnewies, determined the FFs of quarks
and gluons into $\pi^{\pm}$ and $K^{\pm}$ mesons at LO and NLO from genuine
experimental data, adopting two different strategies \cite{4,5}.
In the first analysis \cite{4}, we included in the fit only the data on
$\pi^{\pm}$ and $K^{\pm}$ production taken by the TPC/Two-Gamma Collaboration
\cite{T} at SLAC PEP, with energy $\sqrt{s}= 29$~GeV.
These data combine small statistical errors with fine binning in $x$ and are
thus more constraining than data collected by other experiments in the energy
range from 5.2~GeV to 44~GeV.
Charged-pion and -kaon data at other energies were only used for cross checks.
Furthermore, theoretical predictions obtained using these FFs were compared
with data on the inclusive production of unidentified charged hadrons taken at
the PEP, DESY PETRA, and KEK TRISTAN colliders making an assumption on the
$p/\bar p$ contribution based on information extracted from the TPC data
\cite{T}.
In 1994, when the first analysis \cite{4} was performed, data with identified
quark flavours did not exist.
Therefore, we had to make the assumption that the $s$, $c$, and $b$ ($d$, $c$,
and $b$) quarks fragment into $\pi^\pm$ ($K^\pm$) mesons in the same way.

In our second analysis \cite{5}, we could remove this assumption because new
data on charged-hadron production with partial flavour separation had been
released by the ALEPH Collaboration \cite{7} at LEP1 (see also
Refs.~\cite{A,Laleph,pad}).
Apart from the usual data sample without flavour separation, they also
presented light- and $b$-quark-enriched samples.
Here and in the following, we consider $u$, $d$, and $s$ quarks as light.
In addition, we included in our fits accurate $\pi^\pm$ and $K^\pm$ data from
ALEPH \cite{9} and information on tagged three-jet events from OPAL \cite{10}
at LEP1, which constrained the gluon FFs.
In order to describe the production of charged hadrons in terms of the
$\pi^\pm$ and $K^\pm$ FFs, we needed some information on the fragmentation of
quarks and gluons into protons and antiprotons.
To that end, we introduced a function $f(x)$ which parameterizes the ratio of
the $p/\bar{p}$ and $\pi^\pm$ production cross sections as measured by ALEPH
\cite{9}.

It is clear that this analysis \cite{5} suffered from the lack of specific
data on the fragmentation of tagged quarks and gluons to $\pi^\pm$, $K^\pm$,
and $p/\bar p$ hadrons, rather than just to charged hadrons of unidentified
species.
This lack has been remedied in 1998 by the advent of new data from DELPHI 
\cite{D} at LEP1 and SLD \cite{S} at SLC.
Besides other things, SLD \cite{S} measured the differential production cross
section, as a function of the scaled momentum $x=2p/\sqrt s$, for the
identified hadron species $\pi^\pm$, $K^\pm$, and $p/\bar p$ in $Z$-boson
decays to light, $c$, and $b$ quarks and in such decays without flavour
separation.
Similarly, DELPHI \cite{D} presented $\pi^\pm$, $K^\pm$, and $p/\bar p$
production data, covering the full momentum range up to 45.6~GeV,
discriminating between $Z\to u\bar{u},d\bar{d},s\bar{s}$, $Z\to b\bar{b}$,
and $Z\to q\bar{q}$ events.
There is also new specific information that allows us to better control the
gluon FF.
In fact, new gluon-tagged three-jet data of inclusive charged-hadron
production was released by ALEPH \cite{gA} and OPAL \cite{gO}.
In the OPAL \cite{gO} analysis, the gluon jets were identified using the method
proposed by Gary \cite{17}.
This method is based on rare events of the type
$e^+e^-\to q\bar{q}g_{\mathrm{incl}}$ in which the $q$- and $\bar{q}$-quark
jets appear in the same hemisphere of a multihadronic $e^+e^-$ annihilation
event.
The object $g_{\mathrm{incl}}$, which is taken to be the gluon jet, is
defined by all hadrons observed in the hemisphere opposite to that containing
the $q$- and $\bar{q}$-quark jets.
This method allows for the extraction of the gluon FF with as little
theoretical bias as possible.
It is supposed to be superior to the earlier OPAL \cite{10} measurement of
the gluon FF based on a jet-finding algorithm to define the gluon jets, which
entered the BKK analysis \cite{5}.
The new OPAL \cite{gO} data set comprises more points than the previous one
\cite{10}, and the energy of the gluon jets is now well defined, with a mean
value of $E_{\mathrm{jet}}=40.1$~GeV.

It is the purpose of this work to make full use of all the recent data from
ALEPH \cite{gA}, DELPHI \cite{D}, OPAL \cite{gO}, and SLD \cite{S} in order
to construct new sets of LO and NLO FFs without imposing relations between
the FFs of different flavours, except that we identify the FFs of $u$ and
$d$ quarks into $\pi^\pm$ mesons and those of $u$ and $s$ quarks into $K^\pm$
mesons, as in Refs.~\cite{4,5}.
In the case of protons and antiprotons, we fix the $d$-quark FFs to be half
of the $u$-quark FFs.
For completeness, we also include in our fits the older $\pi^\pm$, $K^\pm$,
and $p/\bar{p}$ production data without flavour separation from ALEPH 
\cite{9}, which also entered the BKK analysis \cite{5}.
In order to be sensitive to the running of $\alpha_s$, {\it i.e.}, to be able
to fit the asymptotic scale parameter,
$\Lambda_{\overline{\mathrm{MS}}}^{(5)}$, we also include the TPC data
\cite{T}, as in Refs.~\cite{4,5}.
There is also precise data on charged-hadron production with flavour
separation from ALEPH \cite{Laleph,pad}, who provided light-, $c$-, and
$b$-quark-enriched samples, and from OPAL \cite{O}, who provided light- and
$b$-quark-enriched samples.
To be on the conservative side, we exclude this data from our fits because the
contribution due to charged hadrons other than $\pi^\pm$, $K^\pm$, and
$p/\bar{p}$ is unknown and may be comparable or larger than the experimental 
errors.
In fact, it turns out that this data yields rather sizeable $\chi^2$ values
relative to the theoretical predictions obtained from our FFs.
We also test our FFs against data from $e^+e^-$ colliders with lower CM
energies, namely, from DESY DORIS \cite{DASP,ARGUS} and PETRA
\cite{Tasse,TASSO}.
Furthermore, we study the longitudinal cross section for the production of
charged hadrons on the $Z$-boson resonance and compare the results with data
from ALEPH \cite{Laleph}, DELPHI \cite{Ldelphi}, and OPAL \cite{Lopal}.

This work is organized as follows.
In Section~2, we summarize the data included in our fits, outline the
theoretical framework, describe our fitting procedure, present our FFs, and
discuss their goodness.
Furthermore, we check our FFs against $e^+e^-$ data which are not included in
our fits.
In Section~3, we investigate for charged-hadron production how well our FFs
satisfy the momentum sum rule, which we did not impose on our fits.
Furthermore, we examine to what extent our FFs also account for the
longitudinal cross section of charged-hadron production on the $Z$-boson
resonance, and how our NLO gluon FF compares with the one of the BKK
set~\cite{5} and with the one recently measured by DELPHI \cite{gD}.
Our conclusions are summarized in Section~4.

\section{Analysis and Results}

\begin{table}[hhh]
\begin{footnotesize}
\renewcommand{\arraystretch}{1.1}
\caption{CM energies, types of data, and $\chi^2_{\mathrm{DF}}$ values
obtained at LO and NLO for various data samples.
Samples not used in the fits are marked by asterisks.}
\begin{center}
\begin{tabular}{c|l|lll|lll}
\hline\hline
 $\sqrt{s}$ [GeV] & Data type & 
 \multicolumn{3}{c|}{\makebox[5cm][c]{$\chi^2_{\mathrm{DF}}$ in NLO}} &
 \multicolumn{3}{c}{\makebox[5cm][c]{$\chi^2_{\mathrm{DF}}$ in LO}} \\
\hline
29.0 & $\sigma^\pi$~(all) & 0.64 \cite{T} & & & 
	      	            0.71 \cite{T} & & \\ 
     & $\sigma^K$~(all) & 1.86 \cite{T} & & & 
	      	          1.40 \cite{T} & & \\ 
     & $\sigma^p$~(all) & 0.79 \cite{T} & & & 
	      	          0.70 \cite{T} & & \\
\hline
91.2 & $\sigma^h$~(all) &
1.28 \cite{D} & 48.1 \cite{pad}$^*$ & 14.4 \cite{O}$^*$ &
1.40 \cite{D} & 53.1 \cite{pad}$^*$ & 15.8 \cite{O}$^*$ \\
 & & 1.32 \cite{S} & & & 1.44 \cite{S} & & \\
 & $\sigma^h$~(uds) &
0.20 \cite{D} & 89.4 \cite{Laleph}$^*$ & 5.10 \cite{O}$^*$ &
0.20 \cite{D} & 92.1 \cite{Laleph}$^*$ & 4.60 \cite{O}$^*$ \\
 & $\sigma^h$~(c) &
 & 80.1 \cite{Laleph}$^*$ & 0.51 \cite{O}$^*$ &
 & 58.9 \cite{Laleph}$^*$ & 0.45 \cite{O}$^*$ \\
 & $\sigma^h$~(b) &
0.43 \cite{D} & 221 \cite{Laleph}$^*$ & 6.63 \cite{O}$^*$ &
0.41 \cite{D} & 220 \cite{Laleph}$^*$ & 6.33 \cite{O}$^*$ \\
\hline
 & $\sigma^\pi$~(all) &
1.28 \cite{9} & 0.58 \cite{D} & 3.09 \cite{S} & 
1.65 \cite{9} & 0.60 \cite{D} & 3.13 \cite{S} \\
 & $\sigma^\pi$~(uds) &
 & 0.72 \cite{D} & 1.87 \cite{S} &
 & 0.73 \cite{D} & 2.17 \cite{S} \\
 & $\sigma^\pi$~(c) &
 & & 1.36 \cite{S} &
 & & 1.16 \cite{S} \\
 & $\sigma^\pi$~(b) &
 & 0.57 \cite{D} & 1.00 \cite{S} &
 & 0.58 \cite{D} & 0.99 \cite{S} \\
\hline
& $\sigma^K$~(all) &
0.30 \cite{9} & 0.86 \cite{D} & 0.44 \cite{S} &
0.32 \cite{9} & 0.79 \cite{D} & 0.45 \cite{S} \\
& $\sigma^K$~(uds) &
 & 0.53 \cite{D} & 0.65 \cite{S} &
 & 0.60 \cite{D} & 0.64 \cite{S} \\
 & $\sigma^K$~(c) &
 & & 2.11 \cite{S} &
 & & 1.90 \cite{S} \\
 & $\sigma^K$~(b) &
 & 0.14 \cite{D} & 1.21 \cite{S} &
 & 0.14 \cite{D} & 1.23 \cite{S} \\
\hline
 & $\sigma^p$~(all) &
0.93 \cite{9} & 0.09 \cite{D} & 0.79 \cite{S} & 
0.80 \cite{9} & 0.06 \cite{D} & 0.70 \cite{S} \\
 & $\sigma^p$~(uds) &
 & 0.11 \cite{D} & 1.29 \cite{S} &
 & 0.14 \cite{D} & 1.28 \cite{S} \\
 & $\sigma^p$~(c) &
 & & 0.92 \cite{S} & 
 & & 0.89 \cite{S} \\
 & $\sigma^p$~(b) &
 & 0.56 \cite{D} & 0.97 \cite{S} &
 & 0.62 \cite{D} & 0.89 \cite{S} \\
\hline
$E_{\mathrm{jet}}$ [GeV] & & & & & & & \\
\hline
26.2 & $D_g^h$ & 1.19 \cite{gA} & & & 1.18 \cite{gA} & & \\
40.1 & $D_g^h$ & 1.03 \cite{gO} & & & 0.90 \cite{gO} & & \\
\hline \hline
\end{tabular}
\end{center}
\end{footnotesize}
\end{table}

We start by summarizing the experimental data which are used for our fits or
for comparisons. We include in our fits the $\pi^\pm$, $K^\pm$, and
$p/\bar p$ data with flavour separation from DELPHI \cite{D} and SLD
\cite{S} and those without flavour separation from ALEPH \cite{9},
DELPHI \cite{D}, SLD \cite{S}, and TPC \cite{T}. Furthermore, we use
the charged-hadron data with flavour separation from DELPHI \cite{D}
and those without flavour separation from SLD \cite{S}. In order to constrain
the gluon FFs, we employ the gluon-tagged three-jet data without hadron
identification from ALEPH \cite{gA} and OPAL \cite{gO}. The
charged-hadron data without and with flavour separation from ALEPH 
\cite{Laleph,pad} and OPAL \cite{O} are only used for comparison. All
these data sets are summarized in Table~1.

The key observable of the experimental analyses and our study is the
scaled-momentum distribution normalized to the total hadronic cross section,
$(1/\sigma_{\mathrm{tot}})d\sigma^h/dx$.
In general, the scaled momentum is defined as $x=2p/\sqrt s$, where $p$ is the
three-momentum of the observed hadron in the CM frame.
In three-jet events, one often defines $x=p/E_{\mathrm{jet}}$, where
$E_{\mathrm{jet}}$ is the gluon-jet energy in the CM frame.
By charge-conjugation invariance, the $e^+e^-$ cross sections for $\pi^+$,
$K^+$, and $p$ production should be the same as those for $\pi^-$, $K^-$, and
$\bar p$ production, respectively.
Therefore, one usually sums over both charges, which is also true for the
production of unidentified charged hadrons.
In turn, also our FFs refer to the sums of particles and antiparticles.

The LO and NLO formalisms for extracting FFs from $e^+e^-$ data were
comprehensively described in Refs.~\cite{4,5} and will, therefore, not be
reviewed here.
The NLO formula for $d\sigma^h/dx$ was originally obtained in Ref.~\cite{3}
and is given in Eq.~(3) of Ref.~\cite{5}.
Deviating from Refs.~\cite{4,5}, we calculate $\sigma_{\mathrm{tot}}$ 
including QCD corrections up to ${\cal O}(\alpha_s^2)$.
We work in the $\overline{\mathrm{MS}}$ renormalization and factorization 
scheme and choose the renormalization scale $\mu$ and the factorization scale
$M_f$ to be $\mu=M_f=\sqrt s$, except for gluon-tagged three-jet events, where
we put $\mu=M_f=2E_{\mathrm{jet}}$. 
For the actual fitting procedure, we use $x$ bins in the interval
$0.1\le x\le1$
and integrate the theoretical functions over the bins width as is done in the
experimental analyses.
The restriction at small $x$ is introduced to exclude events in the
nonperturbative region, where mass effects and nonperturbative
intrinsic-transverse-momentum ($k_T$) effects are important and the underlying
formalism is insufficient.
We parameterize the $x$ dependence of the FFs at the starting scale $Q_0$ as 
\begin{equation}
D_a^h(x,Q_0^2)=Nx^{\alpha}(1-x)^{\beta}.
\label{ansatz} 
\end{equation}
As in Ref.~\cite{5}, we use $Q_0=\sqrt{2}$~GeV for $a=u,d,s,g$,
$Q_0=m(\eta_c)=2.9788$~GeV for $a=c$, and $Q_0=m(\Upsilon)=9.46037$~GeV for
$a=b$.
We impose the conditions
\begin{eqnarray}
D_u^{\pi^\pm}(x,Q_0^2)&=&D_d^{\pi^\pm}(x,Q_0^2),\nonumber\\
D_u^{K^\pm}(x,Q_0^2)&=&D_s^{K^\pm}(x,Q_0^2),\nonumber\\
D_u^{p/\bar{p}}(x,Q_0^2)&=&2D_d^{p/\bar{p}}(x,Q_0^2),
\label{comp}
\end{eqnarray}
which are suggested by the valence quark composition of the respective
hadrons. Equations~(\ref{comp}) are preserved by the $Q^2$ evolution.
The constraint on the $K^\pm$ FFs should be slightly violated by the mass
difference between the $u$ and $s$ quarks \cite{fie}.
In fact, the $s\to K^-$ transition should happen more frequently than the
$\bar u\to K^-$ one because less energy is needed for the creation of a
$u\bar u$ pair from the vacuum than for a $s\bar s$ pair.
However, we shall neglect this subtlety in our analysis because the presently
available experimental information is not rich enough to resolve such minor
differences. For all the other FFs, we treat $N$, $\alpha$, and
$\beta$ as independent fit parameters. In addition, we take the
asymptotic scale parameter $\Lambda_{\overline{\mathrm{MS}}}^{(5)}$,
appropriate for five quark flavours, as a free parameter.
Thus, we have a total of 46 independent fit parameters.
The quality of the fit is measured in terms of the $\chi^2$ value per degree
of freedom, $\chi^2_{\mathrm{DF}}$, for all selected data points.
Using a multidimensional minimization algorithm \cite{18}, we search this
46-dimensional parameter space for the point at which the deviation of the
theoretical prediction from the data becomes minimal.

The $\chi^2_{\mathrm{DF}}$ values achieved for the different data sets
used in our LO and NLO fits may be seen from Table~1. Most of the 
$\chi^2_{\mathrm{DF}}$ values lie around 1 or below, indicating that
the fitted FFs describe all data sets within their respective errors. 
In general, the $\chi^2_{\mathrm{DF}}$ values come out in favour of the
DELPHI \cite{D} data. The overall goodness of the NLO
(LO) fit is given by $\chi^2_{\mathrm{DF}}=$0.98 (0.97). In Table~1, we
have also specified the $\chi^2_{\mathrm{DF}}$ values for those data
sets which we have discussed above, but not included in the fits
\cite{Laleph,pad,O}. Some of these sets come out with much higher
$\chi^2_{\mathrm{DF}}$ values than those included in the fits, which
will be discussed in detail later. The values of the parameters $N$,
$\alpha$, and $\beta$ in Eq.~(\ref{ansatz}) resulting from the LO and
NLO fits are collected in Table~2. They refer to the fragmentation of
the partons $a=u,d,s,c,b,g$ into the hadrons 
$h=\pi^\pm,K^\pm,p/\bar p$, where the sum over particles and
antiparticles is implied.

\begin{table}[hhh]
\begin{small}
\renewcommand{\arraystretch}{1.1}
\caption{\label{pars}Values of $N$, $\alpha$, and $\beta$ in
Eq.~(\protect\ref{ansatz}) resulting from the NLO fit. The numbers given in
parentheses refer to the LO fit.}
\begin{center}
\begin{tabular}{c|l|ccc}
\hline \hline 
Hadron & Flavour & $N$ & $\alpha$ & $\beta$ \\
\hline 
$\pi^\pm$ & $u=d$  & 0.448 (0.546) & $-1.48$ ($-1.47$) & 0.913 (1.02) \\
      & $s$    & 16.6 (22.3) & 0.133 (0.127) & 5.90 (6.14) \\
      & $c$    & 6.17 (8.76) & $-0.536$ ($-0.386$) & 5.60 (5.62) \\
      & $b$    & 0.259 (0.311) & $-1.99$ ($-1.93$) & 3.53 (3.47) \\
      & $g$    & 3.73 (6.05) & $-0.742$ ($-0.714$) & 2.33 (2.92) \\ \hline
$K^\pm$   & $u=s$  & 0.178 (0.259) & $-0.537$ ($-0.619$) & 0.759 (0.859) \\ 
      & $d$    & 4.96 (5.38) & 0.0556 ($-0.00321$) & 2.80 (3.08) \\
      & $c$    & 4.26 (5.18) & $-0.241$ ($-0.178$) & 4.21 (4.30) \\
      & $b$    & 1.32 (1.57) & $-0.884$ ($-0.841$) & 6.15 (6.01) \\
      & $g$    & 0.231 (0.0286) & $-1.36$ ($-2.94$) & 1.80 (2.73) \\ \hline
$p/\bar p$   & $u=2d$ & 1.26 (0.402) & 0.0712 ($-0.860$) & 4.13 (2.80) \\
      & $s$    & 4.01 (4.08) & 0.173 ($-0.0974$) & 5.21 (4.99) \\
      & $c$    & 0.0825 (0.111) & $-1.61$ ($-1.54$) & 2.01 (2.21) \\
      & $b$    & 24.3 (40.1) & 0.579 (0.742) & 12.1 (12.4) \\
      & $g$    & 1.56 (0.740) & 0.0157 ($-0.770$) & 3.58 (7.69) \\
\hline \hline
\end{tabular}
\end{center}
\end{small}
\end{table}

Since we included in our fits high-quality data from two very different
energies, namely 29 and 91.2~GeV, we are sensitive to the running of
$\alpha_s(\mu)$ and are, therefore, able to extract values of 
$\Lambda_{\overline{\mathrm{MS}}}^{(5)}$. We obtain
$\Lambda_{\overline{\mathrm{MS}}}^{(5)}=213{+75\atop-73}$~MeV at NLO 
and $\Lambda_{\overline{\mathrm{MS}}}^{(5)}=88{+34\atop-31}$~MeV at LO.
This corresponds to $\alpha_s(M_Z)=0.1170{+0.005\atop-0.007}$ at NLO and
$\alpha_s(M_Z)=0.1181{+0.006\atop-0.007}$ at LO. Here, the errors are
determined by varying $\Lambda_{\overline{\mathrm{MS}}}^{(5)}$ in such
a way that the total $\chi^2_{\mathrm{DF}}$ is increased by one unit
if all the other fit parameters are kept fixed. They do not include
theoretical errors, which may be estimated through variations of the
overall scale $\mu=M_f$. Our NLO values for
$\Lambda_{\overline{\mathrm{MS}}}^{(5)}$ and $\alpha_s(M_Z)$ 
perfectly agree with those presently quoted by the Particle Data Group as
world averages, $\Lambda_{\overline{\mathrm{MS}}}^{(5)}=212{+25\atop-23}$~MeV
and $\alpha_s(M_Z)=0.1185\pm0.002$, respectively \cite{pdg}.
We note, however, that in Ref.~\cite{pdg} the value of
$\Lambda_{\overline{\mathrm{MS}}}^{(5)}$ is obtained from $\alpha_s(M_Z)$ 
through the next-to-next-to-leading-order formula for $\alpha_s(\mu)$, while
we use the LO and NLO formulas, depending on the order of our analysis, as is
required by consistency.
Since the errors on $\alpha_s(\mu)$ and
$\Lambda_{\overline{\mathrm{MS}}}^{(5)}$ quoted in Ref.~\cite{pdg} result from
an average of different kinds of measurements, they are considerably smaller
than those obtained here from a single type of experiment.
In this context, we remark that, throughout our analysis, we assume
nonperturbative power corrections proportional to $1/Q$ to be negligible in
the energy range relevant for our fits.

The goodness of our LO and NLO fits to the ALEPH \cite{9,gA}, DELPHI \cite{D},
OPAL \cite{gO}, and SLD \cite{S} data may be judged from Figs.~1--5.
In Figs.~1--4, we study the cross section
$(1/\sigma_{\mathrm{tot}})d\sigma^h/dx$ of charged-hadron, 
$\pi^{\pm}$, $K^{\pm}$, and $p/\bar{p}$ production, respectively, at
$\sqrt{s}=91.2$~GeV as a function of $x$. In Fig.~1, the contributions
from all quark flavours and the gluon are included, while Figs.~2--4
refer to the fragmentation of light, $c$, and $b$ quarks, respectively. 
As in the experimental analyses \cite{Laleph,pad,D,S,O}, we have to arrange
for the latter three contributions to add up to the first one.
Therefore, we distribute the gluon contribution among the various quark 
contributions according to their electroweak coupling strengths to the neutral
current. In this way, the gluon radiation off a quark line of a given
flavour is accounted towards the contribution of this quark flavour. 
The theoretical results are compared with the respective data from ALEPH
\cite{9} in Fig.~1, with that from DELPHI \cite{D} in Figs.~1, 2, and 4, and
with that from SLD \cite{S} in Figs.~1--4. We observe that, in all
cases, the various data are mutually consistent with each other and
are nicely described by the LO and NLO fits, which is also reflected
in the relatively small $\chi^2_{\mathrm{DF}}$ values given in 
Table~1. The LO and NLO fits are almost indistinguishable in those
regions of $x$, where the data have small errors. For large $x$,
where the statistical errors are large, the LO and NLO results 
sometimes moderately deviate from each other.

In Fig.~5, we compare the ALEPH \cite{gA} and OPAL \cite{gO} measurements of
the gluon FF in charged-hadron production, which comprises five and twelve
data points, respectively, with our LO and NLO fit results.
The scale choice is $\mu=M_f=2E_{\mathrm{jet}}$, with
$E_{\mathrm{jet}}=26.2$ \cite{gA} and 40.1~GeV \cite{gO}, respectively.
The data are nicely fitted, with $\chi^2_{\mathrm{DF}}$ values of order unity,
as may bee seen from Table~1. By the same token, this implies that the
data are mutually consistent.

As may be seen from Table~1, the ALEPH \cite{Laleph,pad} and OPAL \cite{O}
data on charged-hadron production with and without flavour separation, which
are excluded from our fits, lead to rather sizeable $\chi^2_{\mathrm{DF}}$
values relative to the theoretical predictions based on our FFs.
They are obviously inconsistent with the DELPHI \cite{D} and SLD \cite{S}
data included in our fits.
In particular, the flavour-enriched charged-hadron samples from ALEPH
\cite{Laleph} give $\chi^2_{\mathrm{DF}}$ values of order 100.
The comparison of these ALEPH \cite{Laleph,pad} and OPAL \cite{O} data with
our LO and NLO predictions is also illustrated in Fig.~6.
We should point out that the ALEPH \cite{Laleph,pad,9} data possess the 
following property.
The sum of the three flavour-tagged cross sections of charged-hadron
production \cite{Laleph}, which agrees with the charged-hadron cross section
without flavour separation quoted in Ref.~\cite{pad}, appreciably overshoots
the sum of the $\pi^\pm$, $K^\pm$, and $p/\bar{p}$ cross sections without
flavour separation, which stem from an independent measurement \cite{9}.
The latter three data sets \cite{9} are included in our fits and lead to
quite acceptable $\chi^2_{\mathrm{DF}}$ values, as may be seen from Table~1.
We remark that the charged-hadron data with flavour separation from ALEPH
\cite{Laleph} served as one of the main inputs for the BKK analysis \cite{5}.
We attribute this difference to the contribution from charged hadrons other
than $\pi^\pm$, $K^\pm$, and $p/\bar{p}$ hadrons, which may be comparable or 
larger than the relatively small errors on the charged-hadron data
\cite{Laleph,pad}. We assume that this interpretation is also valid
for the OPAL data \cite{O}.
On the contrary, the DELPHI \cite{D} and SLD \cite{S} analyses are based on
the assumption that the sum of the $\pi^\pm$, $K^\pm$, and $p/\bar{p}$ data
exhaust the full charged-hadron data.
In want of a separate measurement of the excess in cross section due to
high-mass charged hadrons, we decided to proceed as in Refs.~\cite{D,S}.

\begin{table}[hhh]
\begin{small}
\renewcommand{\arraystretch}{1.1}
\caption{CM energies, types of data, and $\chi^2_{\mathrm{DF}}$ values
obtained at LO and NLO for various pre-LEP/SLC data samples not included in 
our fits.}
\begin{center}
\begin{tabular}{c|l|c|c} \hline\hline
 $\sqrt{s}$ [GeV] & Data type & \makebox[3cm][c]{$\chi^2_{\mathrm{DF}}$ in
NLO} & \makebox[3cm][c]{$\chi^2_{\mathrm{DF}}$ in LO} \\
\hline
5.4 & $\sigma^\pi$~(all) & 3.10 \cite{DASP} & 3.02 \cite{DASP} \\ 
    & $\sigma^K$~(all) & 1.80 \cite{DASP} & 2.33 \cite{DASP} \\
\hline
9.98 & $\sigma^\pi$~(all) & 3.27 \cite{ARGUS} & 2.78 \cite{ARGUS} \\ 
     & $\sigma^K$~(all) & 3.21 \cite{ARGUS} & 2.81 \cite{ARGUS} \\
\hline
22.0 & $\sigma^p$~(all) & 1.29 \cite{Tasse} & 1.50 \cite{Tasse} \\
\hline
34.0 & $\sigma^\pi$~(all) & 0.80 \cite{TASSO} & 0.88 \cite{TASSO} \\ 
     & $\sigma^K$~(all) & 0.31 \cite{TASSO} & 0.37 \cite{TASSO} \\ 
     & $\sigma^p$~(all) & 0.58 \cite{TASSO} & 0.46 \cite{TASSO} \\
\hline \hline
\end{tabular}
\end{center}
\end{small}
\end{table}

We now report on comparisons with pre-LEP/SLC data on $\pi^\pm$,
$K^\pm$, and $p\bar p$ production which we did not include in our
fits. Specifically, we consider the $\pi^\pm$ and $K^\pm$ data from
DASP \cite{DASP} ($\sqrt{s}=5.2$~GeV) and ARGUS \cite{ARGUS}
($\sqrt{s}=9.98$~GeV) at DESY  DORIS and from TASSO \cite{TASSO}
($\sqrt{s}=34$~GeV) at PETRA, and the $p/\bar p$ data taken by
TASSO at $\sqrt{s}=22$ \cite{Tasse} and 34~GeV \cite{TASSO}.
There is no separation into flavour-enriched or three-jet samples for
this data. The resulting $\chi^2_{\mathrm{DF}}$ values are summarized
in Table~3. As in the fits, they are evaluated only taking into
account the data points with $x\ge0.1$. The comparison of the
$\pi^\pm$, $K^\pm$, and $p/\bar p$ data with our LO and NLO
predictions is visualized in Figs.~7--9, respectively. For reference,
also the corresponding TPC \cite{T} and SLC \cite{S} data are included
in these comparisons. From Table~3 and Figs.~7--9, we learn that the
DASP \cite{DASP}, ARGUS \cite{ARGUS}, and TASSO \cite{Tasse,TASSO}
data agree quite well with our LO and NLO predictions. This concludes
the description of our fitting procedure and of the quality of the fits.

\section{Applications}

We now investigate if our FFs satisfy the momentum sum rules.
Since a given parton $a$ fragments with 100\% likelihood into some hadron $h$
and momentum is conserved during the fragmentation process, we have 
\begin{equation}
\int_0^1dx\,xD_a^h(x,Q^2)=1,
\label{sumr}
\end{equation}
for any value of $Q^2$.
As is well established experimentally, even for the flavour-enriched data 
samples \cite{S,D}, the sum of the $\pi^\pm$, $K^\pm$, and $p/\bar p$ 
production cross sections agrees, to good approximation, with the one of
charged-hadron production.
Thus, insertion of our $\pi^\pm$, $K^\pm$, and $p/\bar p$ FFs in
Eq.~(\ref{sumr}) exhausts the charged-hadron contribution.
As for the neutral-hadron FFs, which also enter Eq.~(\ref{sumr}), we make the
following assumptions which are suggested by SU(2) flavour symmetry.
As in Ref.~\cite{5}, we approximate
\begin{eqnarray}
D_a^{\pi^0}(x,Q^2) &=& \frac12 D_a^{\pi^\pm}(x,Q^2),\nonumber\\
D_{u,d,s,c,b,g}^{K^0/\bar{K}^0}(x,Q^2) &=& D_{d,u,s,c,b,g}^{K^\pm}(x,Q^2).
\label{ppbar}
\end{eqnarray}
Furthermore, we assume that
\begin{eqnarray}
D_u^{n/\bar{n}}(x,Q^2) &=& \frac12 D_u^{p/\bar{p}}(x,Q^2),\nonumber\\
D_d^{n/\bar{n}}(x,Q^2) &=& 2 D_d^{p/\bar{p}}(x,Q^2),\nonumber\\
D_{s,c,b,g}^{n/\bar{n}}(x,Q^2) &=& D_{s,c,b,g}^{p/\bar{p}}(x,Q^2).
\label{nnbar}
\end{eqnarray}
Here, $D_a^{\pi^\pm}(x,Q^2)$, $D_a^{K^0/\bar{K}^0}(x,Q^2)$, {\it etc.}\ refer
to the sums of the FFs for the individual hadron species.
Equations~(\ref{ppbar}) are supported by $e^+e^-$ data of pion \cite{A,ada}
and kaon \cite{expsum} production, except for very small $x$.
To be on the conservative side, we take the lower limit of integration in
Eq.~(\ref{sumr}) to be 0.05 rather than 0, so that the FF parameterizations
are not used too far outside the $x$ range which was selected for the fits.
This particular choice may be motivated by observing that the $\pi^\pm$,
$K^\pm$, and $p/\bar p$ data sets in Figs.~7--9 are well described by the
theoretical predictions down to $x$ values of order 0.05.  
Since we leave out the $x$ range below 0.05, our results for the left-hand
side of Eq.~(\ref{sumr}) are expected to be somewhat below unity.
A more sophisticated approach \cite{kni} would be to improve the description
of the low-$x$ region by modifying our ansatz for the FFs according to the
so-called modified leading logarithmic approximation (MLLA) \cite{ochs}.

In Table~\ref{sumrule}, we list the results obtained with our LO and NLO FFs
for $Q=\sqrt{2}$, 4, 10, 91, and 200~GeV. The accuracy of our results
is limited by the uncertainties in our assumptions (\ref{ppbar}) and
(\ref{nnbar}), in particular for the lower values of $Q$. In the case
of the $d$ quark, we find values slightly larger than unity at small values
of $Q$, both in LO and NLO.
On the other hand, the $s$-quark values are considerably smaller than unity,
in particular in NLO.
The gluon FF gives values appreciably larger than unity at small values of
$Q$,  both in LO and NLO.
We believe that the imbalance between the $d$- and $s$-quark results is due
to our limited knowledge of the relative magnitude of the valence- and
light-sea-quark contents of the $\pi^\pm$, $K^\pm$, and $p/\bar{p}$ hadrons.
Since the $u$, $d$, and $s$ quarks are always combined in the available data,
there are no individual constraints on their FFs.
In order to trace the source of this imbalance, let us consider the $\pi^\pm$
and $K^\pm$ contributions to the left-hand side of Eq.~(\ref{sumr}) at
$Q_0=\sqrt{2}$~GeV for $a=u,d,s$. In the case of $\pi^\pm$, we have
\begin{eqnarray}
 \int_{0.05}^1 dx\, x D_{u,d}^{\pi^\pm}(x,Q_0^2) &=& 0.41,\nonumber\\
 \int_{0.05}^1 dx\, x D_{s}^{\pi^\pm}(x,Q_0^2) &=& 0.23,
\end{eqnarray}
which nicely exhibits the valence-like character of the $u$ and $d$ quarks in
contrast to the sea-like character of the $s$ quark.
In the case of $K^\pm$, however, we find
\begin{eqnarray}
 \int_{0.05}^1 dx\, x D_{u,s}^{K^\pm}(x,Q_0^2) &=& 0.19,\nonumber\\
 \int_{0.05}^1 dx\, x D_{d}^{K^\pm}(x,Q_0^2) &=& 0.25,
\end{eqnarray}
which shows that the $d$ quark does not behave sea-like, contrary to
expectations. We have thus identified the non-sea-like behaviour of
the $d$ quark in the $K^\pm$ mesons as the source of the violation of
the sum rule in Table~4. 

\begin{table} [ttt]
\renewcommand{\arraystretch}{1.1}
\caption{\label{sumrule}Left-hand side of Eq.~(\protect\ref{sumr}) at NLO for
$a=u,d,s,c,b,g$ and $Q=\protect\sqrt2$, 4, 10, 91, 200~GeV.
We sum over $h=\pi^\pm,\pi^0,K^\pm,K^0,\bar K^0,p,\bar p,n,\bar n$ and
integrate over $0.05\le x\le1$.
The numbers given in parentheses are evaluated with our LO set.}
\begin{center}
\begin{tabular}{c|ccccc} \hline\hline
$a$ & \multicolumn{5}{c}{$Q$ [GeV]} \\
\cline{2-6} & \makebox[2.5cm][c]{$\sqrt2$} & \makebox[2.5cm][c]{4} & 
\makebox[2.5cm][c]{10} & \makebox[2.5cm][c]{91} & \makebox[2.5cm][c]{200} \\
\hline
$u$ & 0.96 (1.13) & 0.96 (1.06) & 0.92 (1.00) & 0.82 (0.87) & 0.78 (0.83) \\
$d$ & 1.05 (1.14) & 1.05 (1.07) & 1.00 (1.01) & 0.89 (0.88) & 0.85 (0.84) \\
$s$ & 0.60 (0.82) & 0.68 (0.82) & 0.67 (0.78) & 0.62 (0.69) & 0.60 (0.67) \\
$c$ & --          & 0.91 (0.97) & 0.88 (0.93) & 0.78 (0.81) & 0.75 (0.78) \\
$b$ & --          & --          & 0.73 (0.80) & 0.65 (0.69) & 0.62 (0.66) \\
$g$ & 1.33 (1.85) & 1.15 (1.32) & 1.00 (1.10) & 0.73 (0.74) & 0.67 (0.67) \\
\hline\hline
\end{tabular}
\end{center}
\end{table}

The analysis of the the longitudinal cross section offers a unique opportunity
to test the gluon FF.
The total cross section of inclusive hadron ($h$) production, which we have
considered so far, may be decomposed into a transverse ($T$) and a
longitudinal ($L$) component,
$d\sigma^h/dx=d\sigma_T^h/dx+d\sigma_L^h/dx$, as indicated in Eq.~(3) of 
Ref.~\cite{5}.
As is well known, at ${\cal O}(\alpha_s^0)$, only transversely polarized 
photons and $Z$ bosons contribute to the cross section.
The longitudinal cross section $d\sigma_L^h/dx$ first appears at
${\cal O}(\alpha_s)$.
Thus, one needs to know it through ${\cal O}(\alpha_s^2)$ in order to test the
gluon FF in NLO.
The ${\cal O}(\alpha_s^2)$ correction to $d\sigma_L^h/dx$ has been calculated
in Refs.~\cite{rn96,Furr}. 
In the case of inclusive hadron production, $d\sigma_L^h/dx$ is given by a
convolution of the longitudinal partonic cross sections with the corresponding
FFs and has the following form \cite{Furr}:
\begin{eqnarray}
\frac{1}{\sigma_{\mathrm{tot}}}\,\frac{d\sigma_L^h}{dx} 
&=&\int_x^1\frac{dz}{z}\left\{\left(\sum_{i=1}^{N_F}Q_{q_i}(s)\right)
\left[\frac{1}{N_F}D_\Sigma^h\left(\frac{x}{z},M_f^2\right)
\frac{1}{\sigma_{\mathrm{tot}}}\,\frac{d\sigma_{L,q}^\Sigma}{dz}
+D_g^h\left(\frac{x}{z},M_f^2\right)
\frac{1}{\sigma_{\mathrm{tot}}}\,\frac{d\sigma_{L,g}}{dz}\right]\right.
\nonumber\\
&&{}+\sum_{i=1}^{N_F}Q_{q_i}(s)D_{(+),i}^h\left(\frac{x}{z},M_f^2\right)
\frac{1}{\sigma_{\mathrm{tot}}}\,\frac{d\sigma_{L,q}^{\mathrm{NS}}}{dz} 
\nonumber\\
&&{}+\left.\sum_{i=1}^{N_F}Q_{q_i}^F(s)
\left[D_{(+),i}^h\left(\frac{x}{z},M_f^2\right)
+\frac{1}{N_F}D_\Sigma^h\left(\frac{x}{z},M_f^2\right)\right]
\frac{1}{\sigma_{\mathrm{tot}}}\,\frac{d\sigma_{L,q}^F}{dz}\right\}.
\label{sigxpol}
\end{eqnarray}
Here, $N_C=3$ and $N_F=5$ denote the numbers of colours and active quark
flavours, respectively, and $Q_{q_i}(s)$ and $Q_{q_i}^F(s)$ represent the
effective electroweak couplings of the quarks to the photon and $Z$ boson.
The latter are defined as
\begin{eqnarray}
Q_{q_i}(s)&=&e_e^2e_{q_i}^2+2e_ev_ee_{q_i}v_{q_i}
\frac{s\left(s-M_Z^2\right)}{\left(s-M_Z^2\right)^2+M_Z^2\Gamma_Z^2}
\nonumber\\
&&{}+\left(v_e^2+a_e^2\right)\left(v_{q_i}^2+a_{q_i}^2\right)
\frac{s^2}{\left(s-M_Z^2\right)^2+M_Z^2\Gamma_Z^2},\nonumber\\
Q_{q_i}^F(s)&=&\left(v_{q_i}^2+a_{q_i}^2\right)a_{q_i}
\left(\sum_{j=1}^{N_F}a_{q_j}\right)
\frac{s\left(s-M_Z^2\right)}{\left(s-M_Z^2\right)^2+M_Z^2\Gamma_Z^2},
\end{eqnarray}
where $M_Z$ and $\Gamma_Z$ are the mass and width of the $Z$ boson, 
respectively, $e_f$ is the electric charge of the fermion $f$ in units of the
positron charge, 
$v_f=\left(T_{3,f}-2e_f\sin^2\theta_w\right)/(2\sin\theta_w$\break
$\cos\theta_w)$ and
$a_f=T_{3,f}/(2\sin\theta_w\cos\theta_w)$ are its vector and axial-vector 
couplings to the $Z$ boson, respectively, $T_{3,f}$ is the third component of
isospin of its left-handed component, and $\theta_w$ is the electroweak mixing
angle.
In the limit $s\ll M_Z^2$, only the QED contribution survives, and we have
$Q_{q_i}\to e_e^2e_{q_i}^2$ and $Q_{q_i}^F\to0$.
The labels $\Sigma$ and NS in Eq.~(\ref{sigxpol}) indicate the flavour singlet
and flavour non-singlet contributions, which are related to the FF 
combinations
\begin{eqnarray}
D_\Sigma^h\left(x,M_f^2\right)&=&\sum_{i=1}^{N_F}\left[
D_{q_i}^h\left(x,M_f^2\right)+D_{\bar q_i}^h\left(x,M_f^2\right)\right],
\nonumber\\
D_{(+),i}^h\left(x,M_f^2\right)&=&
D_{q_i}^h\left(x,M_f^2\right)+D_{\bar q_i}^h\left(x,M_f^2\right)
-\frac{1}{N_F}D_\Sigma^h\left(x,M_f^2\right).
\end{eqnarray}
The contributions proportional to the colour factor $C_FT_R=4/8$ are labeled
$F$, which is to suggest that they vanish in QED due to Furry's theorem.
They only arise from $Z$-boson exchange, provided one does not sum over all
quark flavours belonging to one fermion generation.
Furthermore, they are devoid of collinear singularities, so that mass
factorization is not needed.
We neglect these contributions, since they amount to less than 1\% over the
whole $x$ range \cite{Furr}.
The partonic cross sections $d\sigma_{L,q}^\Sigma/dz$,
$d\sigma_{L,g}^\Sigma/dz$, and $d\sigma_{L,q}^{\mathrm{NS}}/dz$ through
${\cal O}(\alpha_s^2)$ may be found in Refs.~\cite{rn96,Furr}.

ALEPH \cite{Laleph}, DELPHI \cite{Ldelphi}, and OPAL \cite{Lopal} have 
measured $d\sigma_L^h/dx$ for charged-hadron production as a function of $x$
without quark flavour separation.
In addition, DELPHI \cite{Ldelphi} has also presented light- and 
$b$-quark-enriched samples. In Fig.~10, these data are compared with
the LO and NLO evaluations of Eq.~(\ref{sigxpol}) with our FFs.
As in our fits to data of $d\sigma^h/dx$, we choose the scales to be equal to
the CM energy $\sqrt{s}=91.2$~GeV, where the data have been taken.
Although only data with $x\ge0.1$ has been included in our fits,
$d\sigma_L^h/dx$ is also well described down to $x=0.01$.
The $\chi^2_{\mathrm{DF}}$ values, evaluated from the data points with 
$x\ge0.1$, are collected in Table~5.
The $\chi^2_{\mathrm{DF}}$ values for the LO predictions are in average more
than a factor of two larger than those for the NLO predictions.
This is not surprising, since it was already found in Ref.~\cite{5} that the
LO result falls short of the data by a factor of two.
At that time, the common scale had to be reduced to 20~GeV in order to obtain
agreement. We observe from Fig.~10 and Table~5 that this is no longer
necessary for the NLO results. Obviously, the ${\cal O}(\alpha_s^2)$
corrections \cite{rn96,Furr} are sufficient to produce a $K$ factor that
brings our predictions into good agreement with the data.
Good agreement between the NLO prediction of $d\sigma_L^h/dx$ and the ALEPH
\cite{Laleph} and OPAL \cite{Lopal} data was already found by Rijken and van
Neerven \cite{rn96}, who employed the BKK gluon FF \cite{4}.

\begin{table}
\begin{small}
\renewcommand{\arraystretch}{1.1}
\caption{CM energies, types of data, and $\chi^2_{\mathrm{DF}}$ values
obtained at LO and NLO for the longitudinal cross section.}
\begin{center}
\begin{tabular}{c|l|lll|lll}
\hline\hline
 $\sqrt{s}$ [GeV] & Data type & 
 \multicolumn{3}{c|}{\makebox[5cm][c]{$\chi^2_{\mathrm{DF}}$ in NLO}} &
 \multicolumn{3}{c}{\makebox[5cm][c]{$\chi^2_{\mathrm{DF}}$ in LO}} \\
\hline
91.2 & $\sigma_L^h$~(all) &
1.36 \cite{Laleph} & 1.74 \cite{Ldelphi} & 0.49 \cite{Lopal} &
11.0 \cite{Laleph} & 1.27 \cite{Ldelphi} & 7.94 \cite{Lopal} \\ 
 & $\sigma_L^h$~(uds) &
 & 7.98 \cite{Ldelphi} & & 
 & 1.05 \cite{Ldelphi} & \\
 & $\sigma_L^h$~(b) &
 & 0.51 \cite{Ldelphi} & &
 & 0.80 \cite{Ldelphi} & \\
\hline \hline
\end{tabular}
\end{center}
\end{small}
\end{table}

The comparisons presented in Fig.~10 provide us with a very useful check of
the gluon FF in the low-$x$ range, where the data have small errors.
In particular, the good agreement in the case of flavour separation nicely
confirms the relative importance of the light- and $b$-quark FFs relative to
the gluon FF. Since the $\chi^2_{\mathrm{DF}}$ values obtained at NLO
for the longitudinal cross section are comparable to those achieved in
our NLO fit, we do not expect that the inclusion of the information on
the longitudinal cross section in our NLO fit would significantly
change the outcome. 

As is well known, the gluon FF only enters the prediction for $d\sigma^h/dx$
at ${\cal O}(\alpha_s)$, while at ${\cal O}(\alpha_s^0)$ it only contributes
indirectly via the $Q^2$ evolution.
Therefore, we have used the ALEPH \cite{gA}
and OPAL \cite{gO} data on gluon-tagged three-jet events in order to
constrain it. These data were already compared with the fitted gluon
FF in Fig.~5. It is of interest to compare this gluon FF with the BKK
\cite{5} one and the one recently extracted by DELPHI \cite{gD} from
their three-jet events for scales in the range $6.5\le M_f\le28$~GeV.
In Ref.~\cite{gD}, a parameterization, valid for this $M_f$ range
and for $0.15\le x\le0.75$, was generated adopting an ansatz similar to the
one introduced in Ref.~\cite{5}. In Fig.~11, the $x$ dependences of
these three gluon FFs are studied for $M_f=10$, 52.4, 80.2, and 200~GeV.
The results for $M_f=52.4$ and 80.2~GeV are compared with the ALEPH \cite{gA}
and OPAL \cite{gO} gluon-tagged three-jet data, which are included in our
fits. From Fig.~11 the following observations can be made.
Our new gluon FF is rather similar to the BKK one, especially for
larger values of $M_f$. At $M_f=80.2$~GeV, the DELPHI gluon FF is
somewhat steeper than the new and BKK ones, except for large $x$
values, and it agrees slightly better with the OPAL \cite{gO} data.
On the other hand, our new gluon FF agrees best with the ALEPH \cite{gA} data
at $M_f=52.4$~GeV.
At $M_f=10$~GeV, the DELPHI gluon FF has a flatter $x$ dependence than the 
other ones and overshoots them, while at $M_f=200$~GeV the situation is
similar to the case of $M_f=80.2$~GeV.

\section{Conclusions}

The LEP1 and SLC experiments have provided us with a wealth of high-precision
experimental information on how partons fragment into low-mass charged
hadrons \cite{A,7,Laleph,pad,9,10,D,S,gA,gO,O,Ldelphi,Lopal,gD,ada}.
The data partly come as light-, $c$-, and $b$-quark-enriched samples without
\cite{Laleph,pad,O} or with identified final-state hadrons ($\pi^\pm$,
$K^\pm$, and $p/\bar p$) \cite{D,S} or as gluon-tagged three-jet samples 
without hadron identification \cite{gA,gO,gD}.
This new situation motivated us to update, refine, and extend the BKK analysis
\cite{5} by generating new LO and NLO sets of $\pi^\pm$, $K^\pm$, and
$p/\bar p$ FFs.
The $x$ distributions of the resulting FFs at their starting scales $Q_0$ are
given by Eq.~(\ref{ansatz}) with the parameters listed in Table~2.
The $Q^2$ evolution is determined by the timelike Altarelli-Parisi equations 
in the respective orders, LO or NLO, which are summarized in the Appendix of
Ref.~\cite{bin}.\footnote{%
There is an obvious typographical error in the published version of
Ref.~\cite{bin}, which was absent in the preprint version thereof.
In the line before the last of Eq.~(17), $\ln\ln(1-x)$ should be replaced by
$\ln(1-x)$.}
The evolution procedure in Mellin space is described in Ref.~\cite{5}.
A FORTRAN subroutine which returns the values of the $D_a^h(x,Q^2)$ functions
for given values of $x$ and $Q^2$ may be downloaded from the URL
{\tt http://www.desy.de/\~{}poetter/kkp.html} or obtained upon
request from the authors.

By also including in our fits $\pi^\pm$, $K^\pm$, and $p/\bar p$ data (without
flavour separation) from PEP \cite{T}, with CM energy $\sqrt s=29$~GeV, we
obtained a handle on the scaling violation in the fragmentation process, which
allowed us to extract LO and NLO values of $\alpha_s(M_Z)$.
We found $\alpha_s(M_Z)=0.1181{+0.006\atop-0.007}$ at LO and
$\alpha_s(M_Z)=0.1170{+0.005\atop-0.007}$ at NLO, where the errors are 
experimental.
These results are in perfect agreement with what the Particle Data Group 
currently quotes as the world average, $\alpha_s(M_Z)=0.1185\pm0.002$
\cite{pdg}.

Our strategy was to only include in our fits LEP1 and SLC data with both
flavour separation and hadron identification \cite{S,D}, gluon-tagged
three-jet samples with a fixed gluon-jet energy \cite{gA,gO}, and the 
$\pi^\pm$, $K^\pm$, and $p/\bar p$ data sets from the pre-LEP1/SLC era with
the highest statistics and the finest binning in $x$ \cite{T}.
Other data served us for cross checks.
In particular, we probed the scaling violation through comparisons with
$\pi^\pm$, $K^\pm$, and $p/\bar p$ data from DORIS and PETRA, with CM energies
between 5.4 and 34~GeV \cite{DASP,ARGUS,Tasse,TASSO}.
Furthermore, we tested the gluon FF, which enters the unpolarized cross
section only at NLO, by comparing our predictions for the longitudinal cross
section, where it already enters at LO, with available data
\cite{Laleph,Ldelphi,Lopal}.
Finally, we directly compared our gluon FF with the one recently measured by
DELPHI in three-jet production with gluon identification as a function of $x$
at various scales $Q^2$ \cite{gD}.
All these comparisons led to rather encouraging results.
We also verified that our FFs satisfy reasonably well the momentum sum
rules, which we did not impose as constraints on our fits.
All these cross checks make us believe that our FFs should allow for a 
reliable description of inclusive charged-hadron production in all kinds of
high-energy-collision experiments over wide ranges of $x$ and $Q^2$.
The very-high-$Q^2$ regime will be accessible with the CERN LHC and a future
$e^+e^-$ linear collider.
Our FFs are bound to break down at very low values of $x$, where it becomes
necessary to modify the ansatz~(\ref{ansatz}) according to the MLLA 
\cite{ochs}.
This is beyond the scope of the present analysis, but should eventually be 
investigated \cite{kni}.

\vspace{1cm}
\noindent
{\bf Note added}
\smallskip

\noindent
After the submission of this paper, a preprint \cite{kre} appeared which also 
presents NLO FFs for $\pi^\pm$, $K^\pm$, and charged hadrons.
These FFs are fitted to the ALEPH \cite{Laleph,pad}, SLD \cite{S}, and TPC
\cite{T} data assuming certain power laws in the light-quark sector.

\vspace{1cm}
\noindent
{\bf Acknowledgements}
\smallskip

\noindent
The II. Institut f\"ur Theoretische Physik is supported in part by the
Deutsche Forschungsgemeinschaft through Grant No.\ KN~365/1-1, by the
Bundesministerium f\"ur Bildung und Forschung through Grant No.\ 05~HT9GUA~3,
and by the European Commission through the Research Training Network
{\it Quantum Chromodynamics and the Deep Structure of Elementary Particles}
under Contract No.\ ERBFMRX-CT98-0194.

\newpage

\begin{figure}[hhh]
  \unitlength1mm
  \begin{picture}(122,160)
    \put(3,0){\epsfig{file=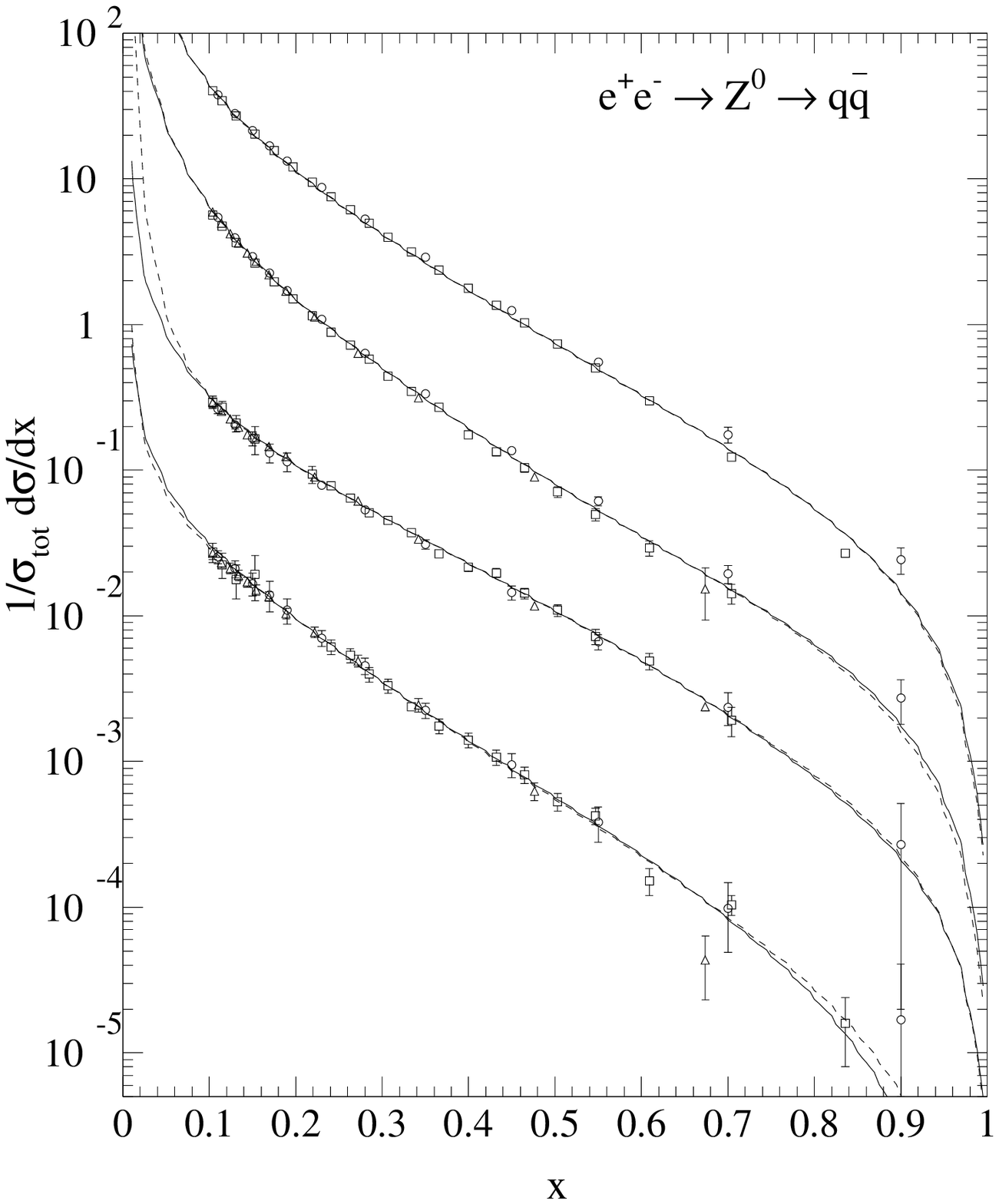,width=16cm}}
  \end{picture}
\caption{Normalized differential cross section of inclusive hadron production
at $\protect\sqrt{s}=91.2$~GeV as a function of $x$.
The LO (dashed lines) and NLO (solid lines) fit results are compared with data
from ALEPH \protect\cite{9} (triangles), DELPHI \protect\cite{D} (circles),
and SLD \protect\cite{S} (squares).
The upmost, second, third, and lowest curves refer to charged hadrons,
$\pi^\pm$, $K^\pm$, and $p/\bar{p}$, respectively.
Each pair of curves is rescaled relative to the nearest upper one by a factor
of 1/5.}
\end{figure}

\newpage

\begin{figure}[hhh]
  \unitlength1mm
  \begin{picture}(122,160)
    \put(3,0){\epsfig{file=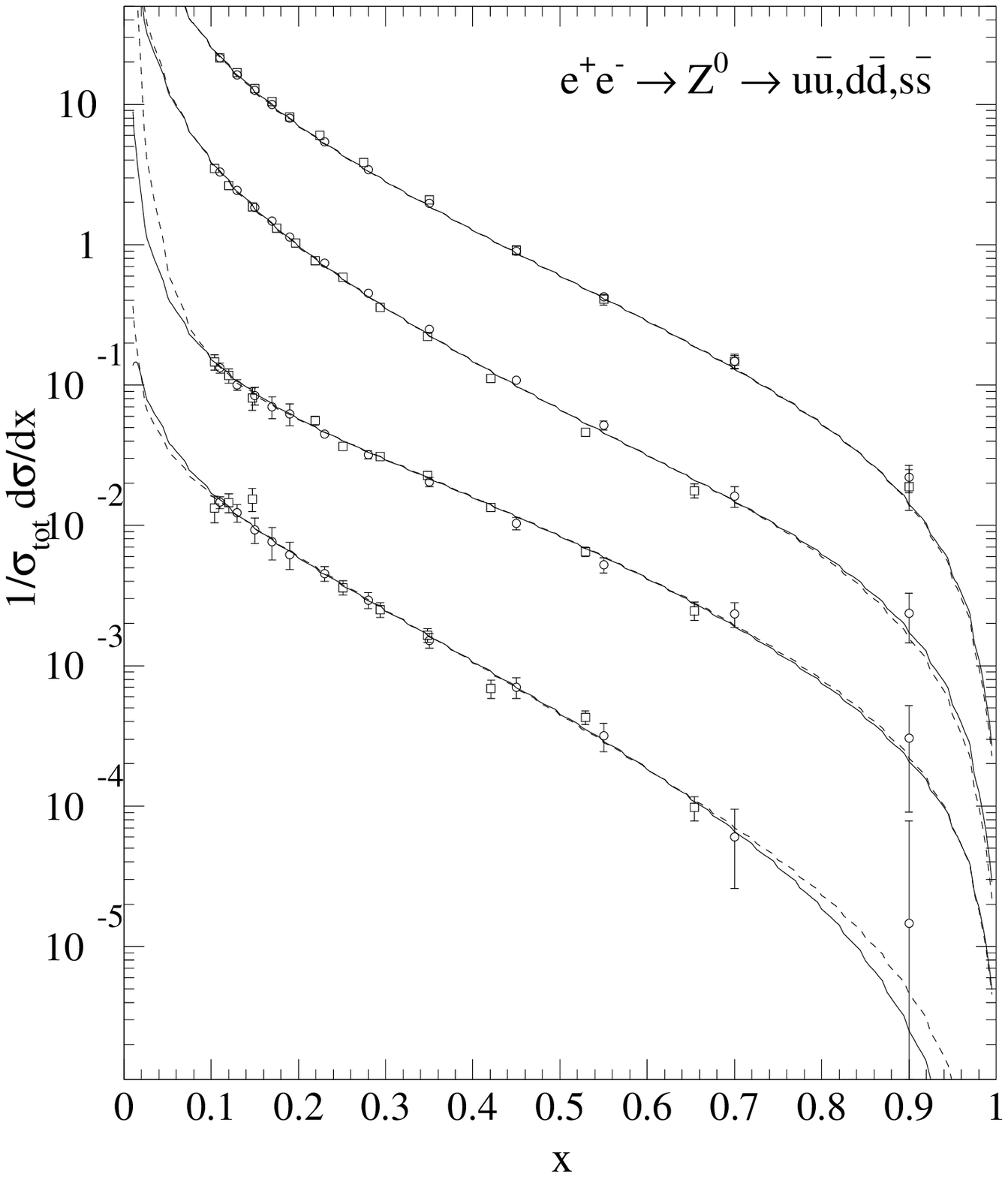,width=16cm}}
  \end{picture}
\caption{Same as in Fig.~1, but for the light-quark-enriched samples from
DELPHI \protect\cite{D} (circles) and SLD \protect\cite{S} (squares).}
\end{figure}

\newpage

\begin{figure}[hhh]
  \unitlength1mm
  \begin{picture}(122,160)
    \put(3,0){\epsfig{file=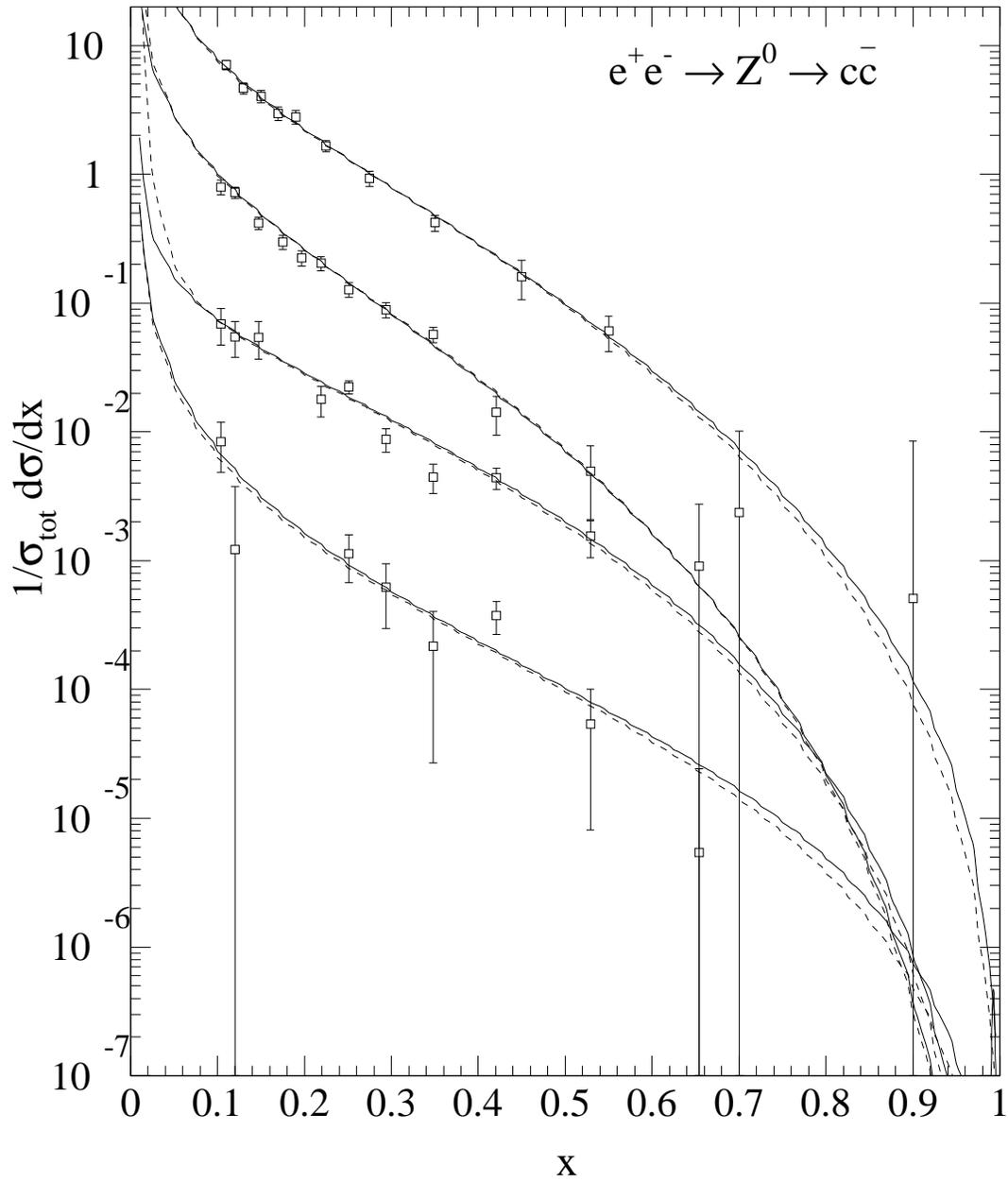,width=16cm}}
  \end{picture}
\caption{Same as in Fig.~1, but for the $c$-quark-enriched samples from SLD
\protect\cite{S} (squares). The last two points on the right belong to
the charged-hadron sample.} 
\end{figure}

\newpage

\begin{figure}[hhh]
  \unitlength1mm
  \begin{picture}(122,160)
    \put(3,0){\epsfig{file=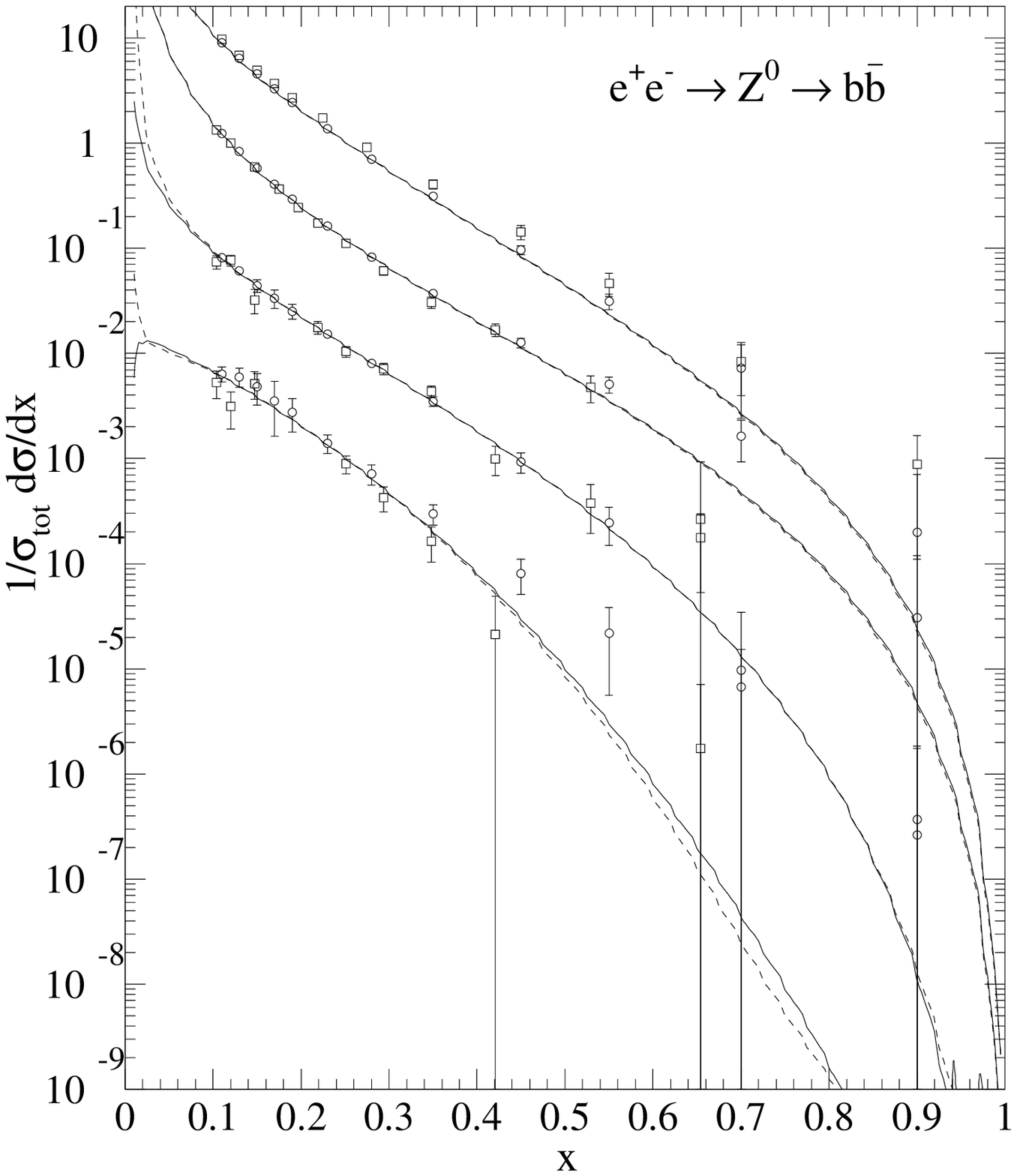,width=16cm}}
  \end{picture}
\caption{Same as in Fig.~1, but for the $b$-quark-enriched samples from
DELPHI \protect\cite{D} (circles) and SLD \protect\cite{S} (squares).}
\end{figure}

\newpage

\begin{figure}[hhh]
  \unitlength1mm
  \begin{picture}(122,160)
    \put(3,0){\epsfig{file=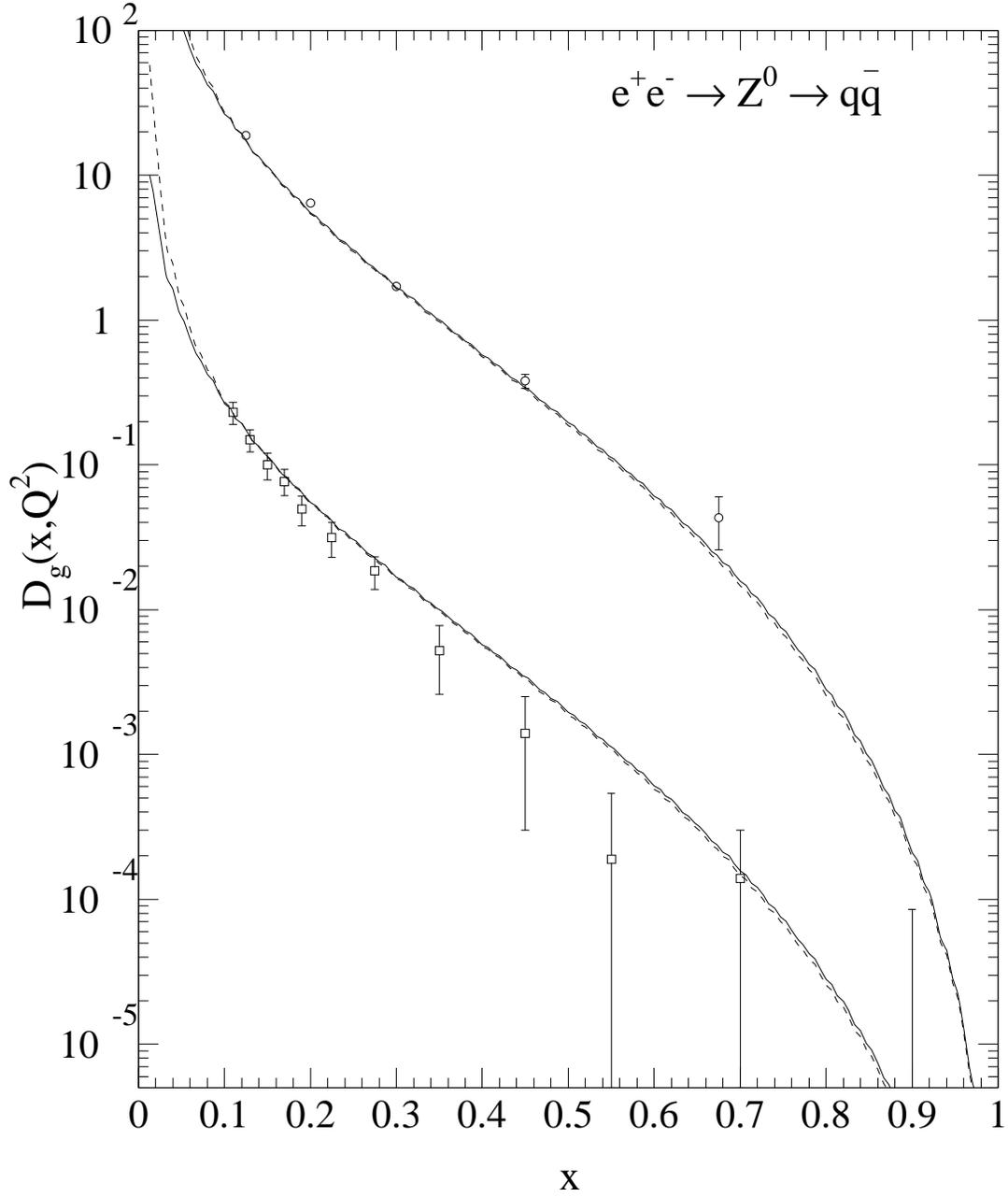,width=16cm}}
  \end{picture}
\caption{Gluon FF for charged-hadron production as a function of $x$ at
$M_f=52.4$ and 80.2~GeV.
The LO (dashed lines) and NLO (solid lines) predictions are compared with 
three-jet data from ALEPH \protect\cite{gA}, with $E_{\mathrm{jet}}=26.2$~GeV,
(upper curves) and from OPAL \protect\cite{gO}, with
$E_{\mathrm{jet}}=40.1$~GeV (lower curves). The OPAL data and the
pertinent predictions are rescaled by a factor of 1/100.}  
\end{figure}

\newpage

\begin{figure}[hhh]
  \unitlength1mm
  \begin{picture}(122,160)
    \put(3,0){\epsfig{file=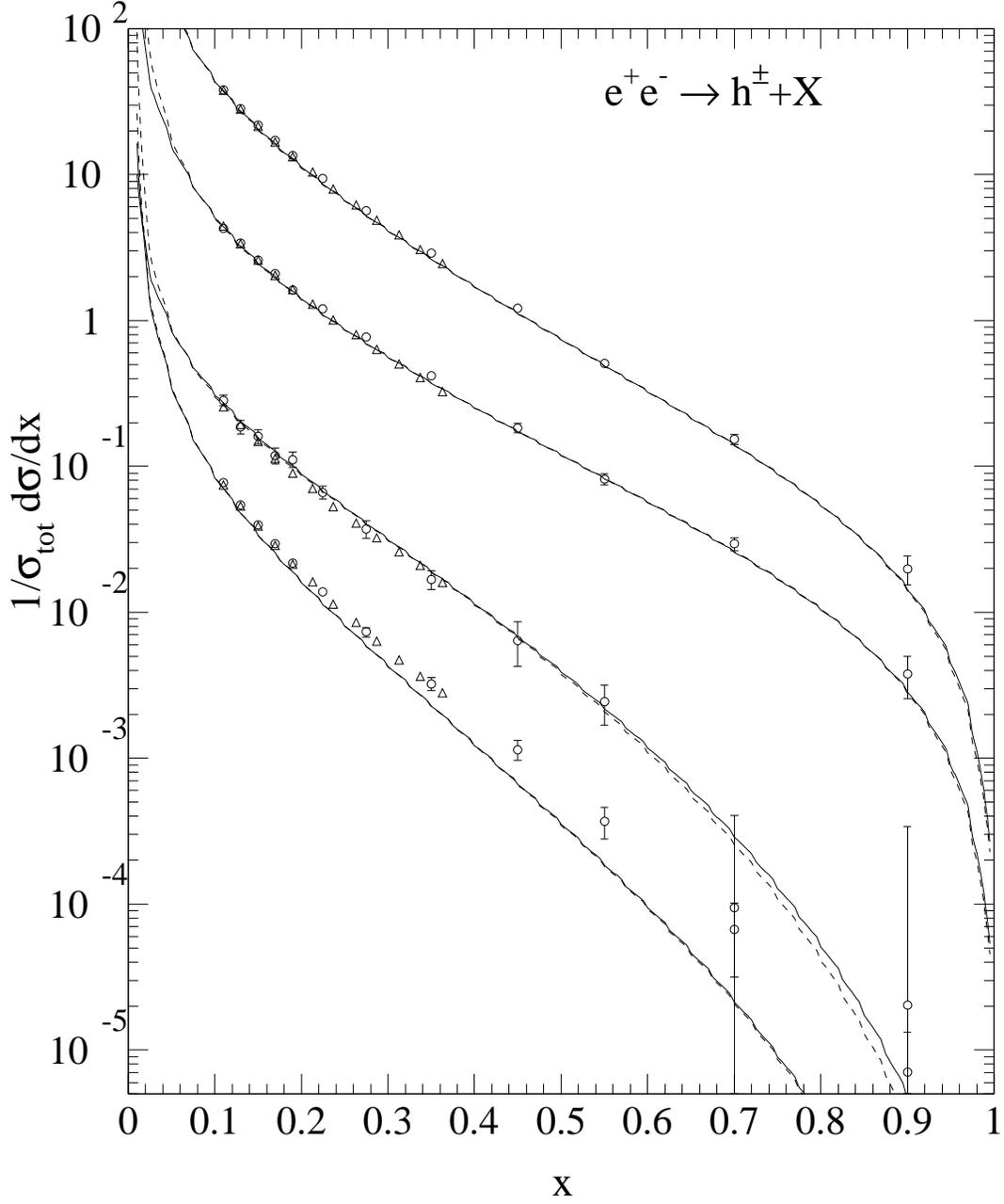,width=16cm}}
  \end{picture}
\caption{Normalized differential cross section of inclusive charged-hadron
production at $\protect\sqrt{s}=91.2$~GeV as a function of $x$.
The LO (dashed lines) and NLO (solid lines) fit results are compared with data
from ALEPH \protect\cite{Laleph,pad} (triangles) and OPAL \protect\cite{O}
(circles).
The upmost, second, third, and lowest curves refer to the full,
light-quark-enriched, $c$-quark-enriched, and $b$-quark-enriched samples,
respectively. Each pair of curves is rescaled relative to the nearest
upper one by a factor of 1/5.}
\end{figure}

\newpage

\begin{figure}[hhh]
  \unitlength1mm
  \begin{picture}(122,160)
    \put(3,0){\epsfig{file=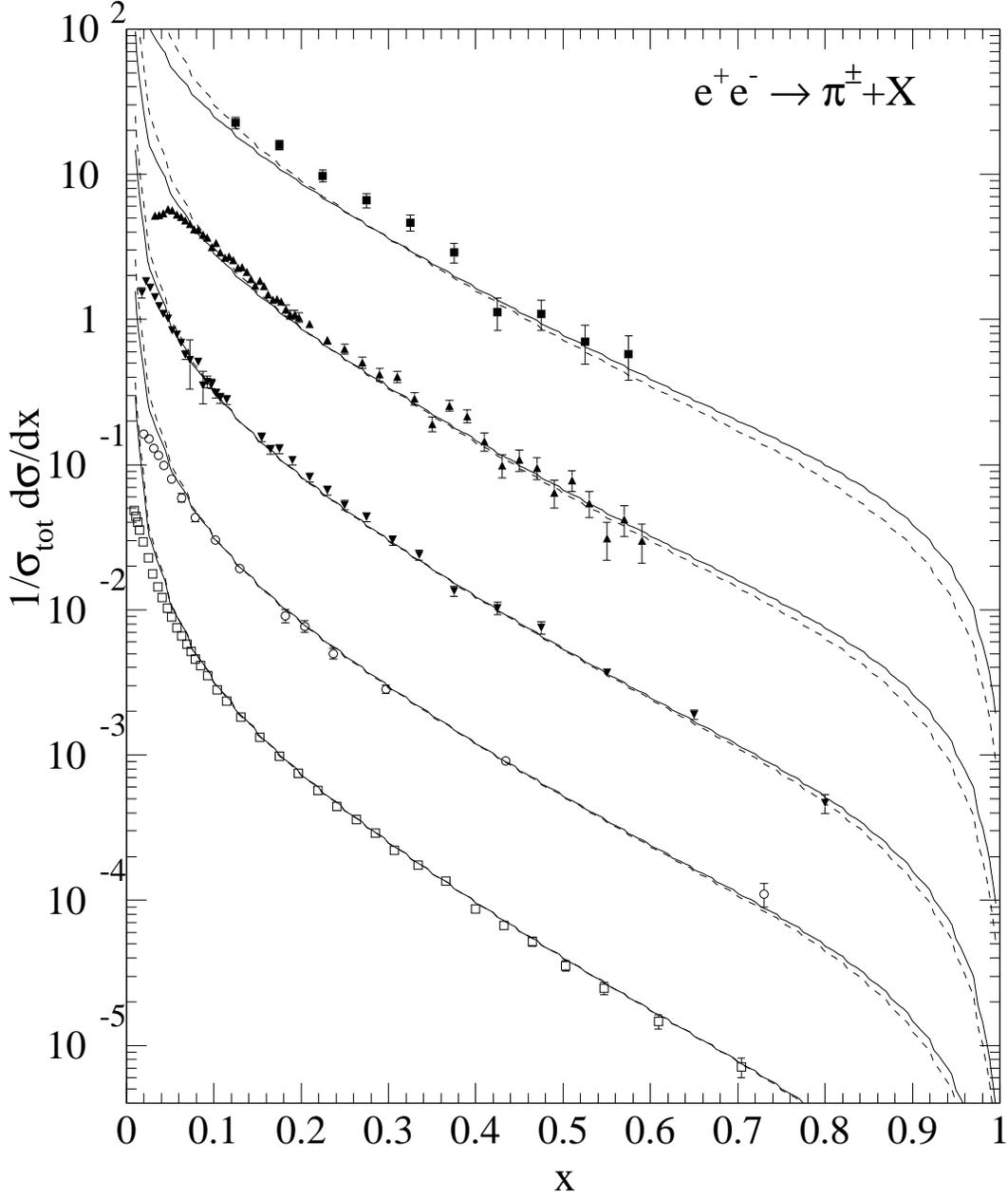,width=16cm}}
  \end{picture}
\caption{Normalized differential cross section of inclusive $\pi^\pm$
production as a function of $x$ at $\protect\sqrt{s}=5.2$, 9.98, 29, 34,
and 91.2~GeV.
The LO (dashed lines) and NLO (solid lines) predictions are compared with data
from DASP \protect\cite{DASP}, ARGUS \protect\cite{ARGUS},
TPC \protect\cite{T}, TASSO \protect\cite{TASSO}, and SLD \protect\cite{S}.
Upper curves correspond to lower energies. Each pair of curves is
rescaled relative to the nearest upper one by a factor of 1/10.}
\end{figure}

\newpage

\begin{figure}[hhh]
  \unitlength1mm
  \begin{picture}(122,160)
    \put(3,0){\epsfig{file=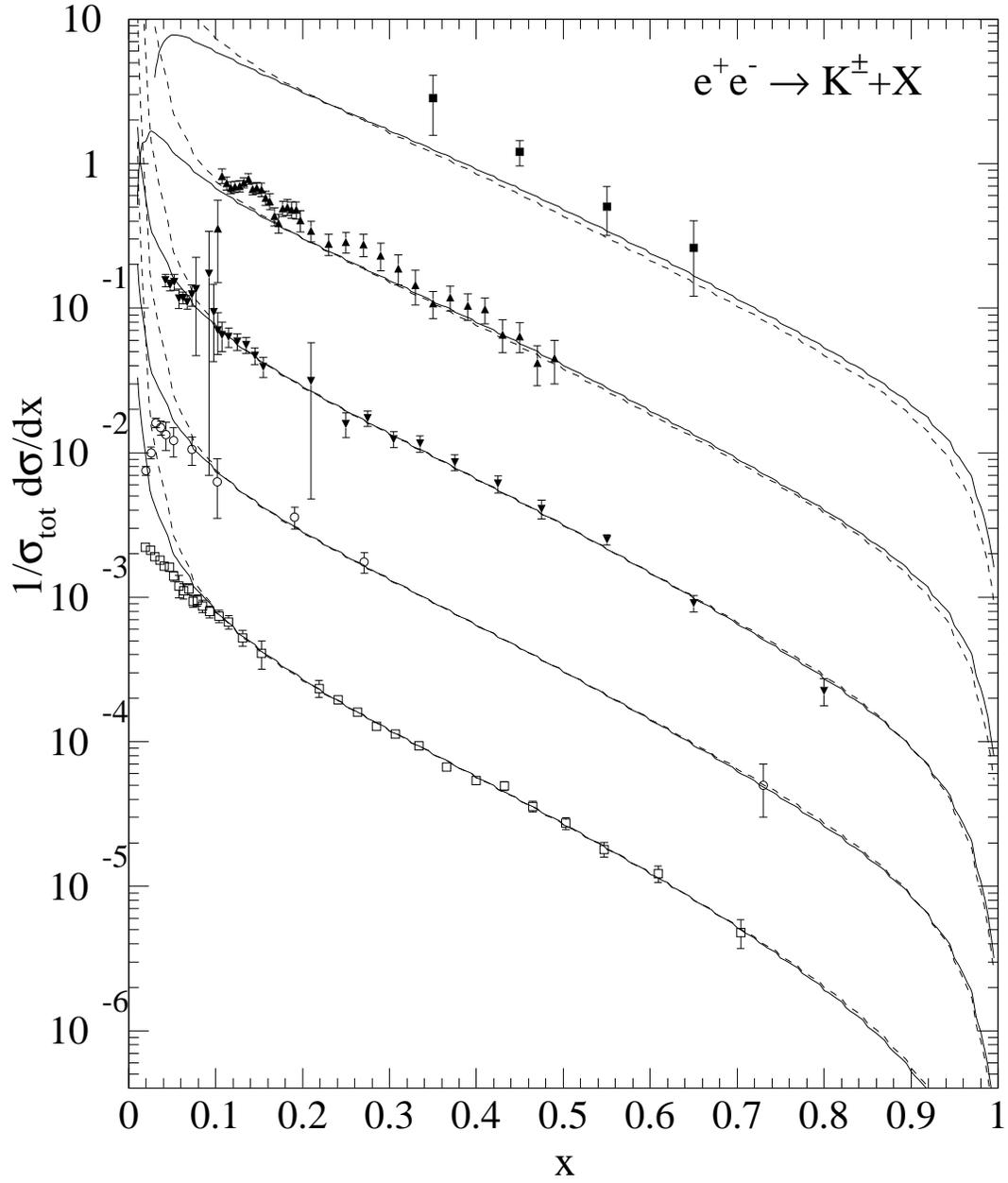,width=16cm}}
  \end{picture}
 \caption{Same as in Fig.~7, but for $K^\pm$ mesons.}
\end{figure}

\newpage

\begin{figure}[hhh]
  \unitlength1mm
  \begin{picture}(122,160)
    \put(3,0){\epsfig{file=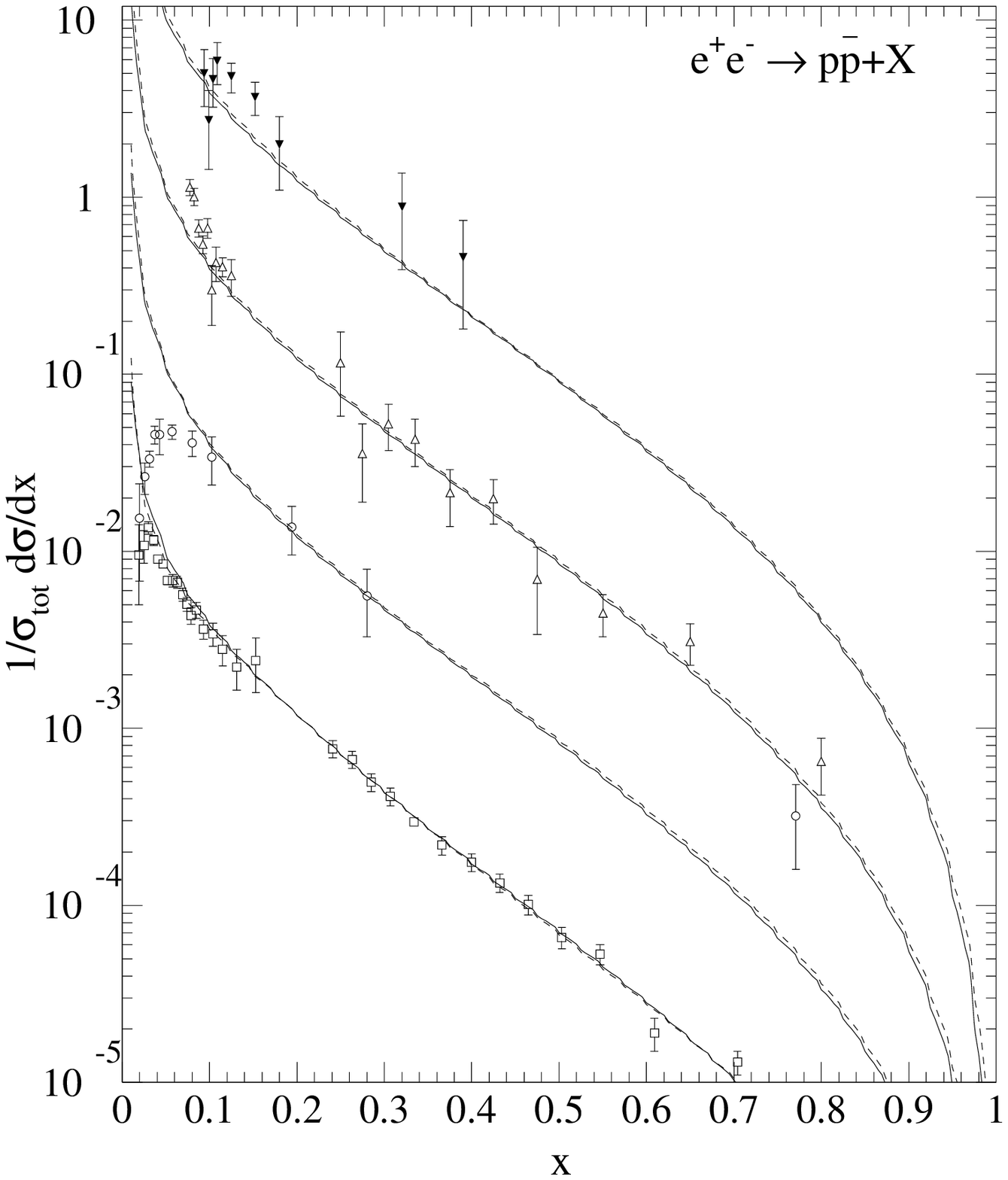,width=16cm}}
  \end{picture}
\caption{Normalized differential cross section of inclusive $p/\bar p$
production as a function of $x$ at $\protect\sqrt{s}=22$, 29, 34, and 
91.2~GeV.
The LO (dashed lines) and NLO (solid lines) predictions are compared with data
from TASSO (with $\protect\sqrt{s}=22$ \protect\cite{Tasse} and 34~GeV
\protect\cite{TASSO}), TPC \protect\cite{T}, and SLD \protect\cite{S}.
Upper curves correspond to lower energies. Each pair of curves is
rescaled relative to the nearest upper one by a factor of 1/10.}
\end{figure}

\newpage

\begin{figure}[hhh]
  \unitlength1mm
  \begin{picture}(122,160)
    \put(3,0){\epsfig{file=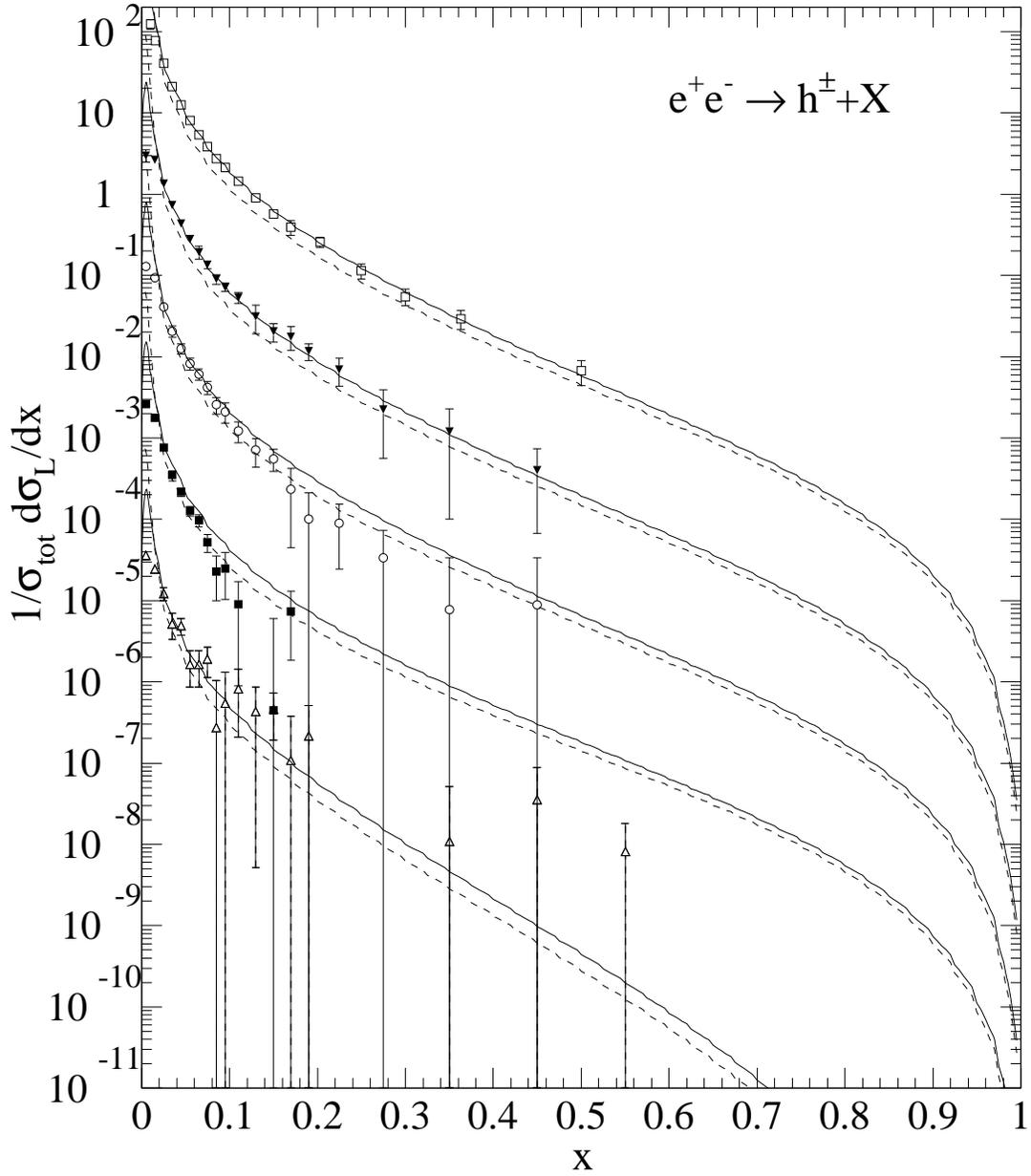,width=16cm}}
  \end{picture}
\caption{Normalized differential longitudinal cross section of inclusive
charged-hadron production as a function of $x$ at $\protect\sqrt{s}=91.2$~GeV.
The LO (dashed lines) and NLO (solid lines) predictions are compared with data
from ALEPH \protect\cite{Laleph}, OPAL \protect\cite{Lopal}, and DELPHI
\protect\cite{Ldelphi} without flavour separation and with light- and
$b$-quark-enriched samples from DELPHI \protect\cite{Ldelphi} (in this order
from top to bottom). Each pair of curves is rescaled relative to the
nearest upper one by a factor of 1/30.} 
\end{figure}

\newpage

\begin{figure}[hhh]
  \unitlength1mm
  \begin{picture}(122,160)
    \put(3,0){\epsfig{file=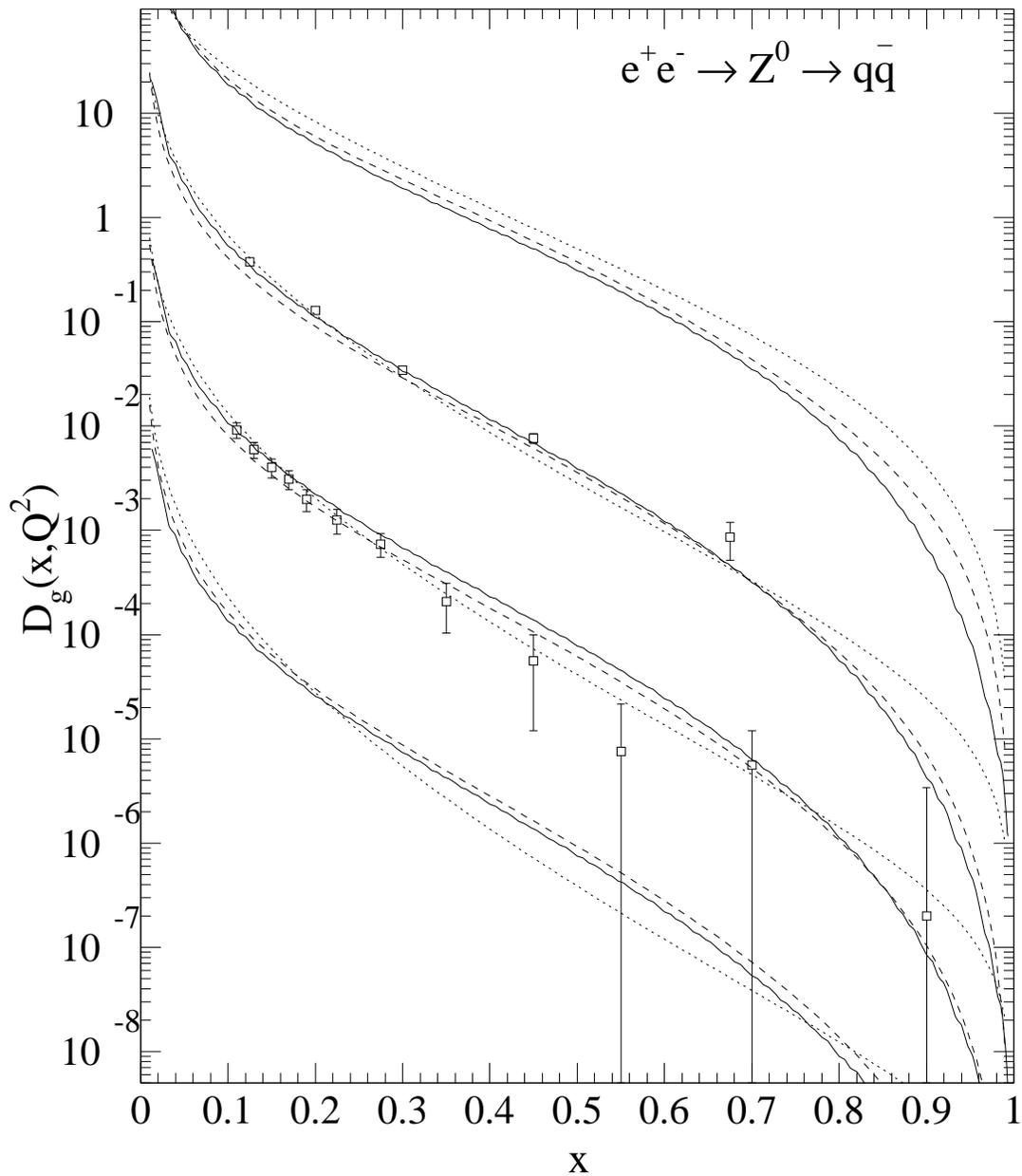,width=16cm}}
  \end{picture}
\caption{Gluon FFs for charged-hadron production as functions of $x$ at
$M_f=10$, 52.4, 80.2, and 200~GeV from the DELPHI \protect\cite{gO} (dotted
lines), BKK NLO \protect\cite{5} (dashed lines), and new NLO (solid lines)
sets. The predictions for $M_f=52.4$ and 80.2~GeV are compared with
the gluon-tagged three-jet data from ALEPH \protect\cite{gA} and OPAL
\protect\cite{gO}, respectively. Upper curves correspond to lower energies.
Each set of curves is rescaled relative to the nearest upper one by a factor
of 1/50.}
\end{figure}

\end{document}